\documentclass[]{aa} 

\usepackage{epsfig}

\newcommand{\pl}{\partial}
\newcommand{\inta}{\int_{-i\infty}^{+i\infty}} 
\newcommand{\beq}{\begin{equation}} 
\newcommand{\eeq}{\end{equation}} 
\newcommand{\beqa}{\begin{eqnarray}} 
\newcommand{\eeqa}{\end{eqnarray}} 
\newcommand{\bea}{\begin{array}} 
\newcommand{\ea}{\end{array}} 
\newcommand{\lag}{\langle} 
\newcommand{\rag}{\rangle}
\newcommand{\Om}{\Omega_{\rm m0}}
\newcommand{\Ode}{\Omega_{\rm de0}}
\newcommand{\OL}{\Omega_{\Lambda 0}}
\newcommand{\Omz}{\Omega_{\rm m}}
\newcommand{\Odez}{\Omega_{\rm de}}
\newcommand{\rhobm}{\overline{\rho}_{\rm m}}
\newcommand{\rhom}{\rho_{\rm m}}
\newcommand{\rhobde}{\overline{\rho}_{\rm de}}
\newcommand{\bx}{{\bf x}}
\newcommand{\cG}{{\cal G}}
\newcommand{\dd}{{\rm d}}
\newcommand{\tdelta}{\tilde{\delta}}
\newcommand{\bk}{{\bf k}}
\newcommand{\ii}{{\rm i}}
\newcommand{\zi}{z_{\rm i}}
\newcommand{\deltaLi}{\delta_{L\rm i}}
\newcommand{\cF}{{\cal F}}
\newcommand{\cP}{{\cal P}}
\newcommand{\cD}{{\cal D}}
\newcommand{\cS}{{\cal S}}
\newcommand{\bq}{{\bf q}}
\newcommand{\bs}{{\bf s}}
\newcommand{\nb}{\overline{n}}
\newcommand{\tW}{\tilde{W}}
\newcommand{\rhob}{\overline{\rho}}

\begin{document}

\title{Mass functions and bias of dark matter halos}    
\author{P. Valageas}   
\institute{Institut de Physique Th\'eorique, CEA Saclay, 91191 Gif-sur-Yvette, 
France}  
\date{Received / Accepted } 
 
\abstract
{}
{We revisit the study of the mass functions and the bias of dark matter halos.}
{Focusing on the limit of rare massive halos, we point out that exact analytical
results can be obtained for the large-mass tail of the halo mass function.
This is most easily seen from a steepest-descent approach, that becomes
asymptotically exact for rare events. 
We also revisit the traditional derivation of the bias of massive halos,
associated with overdense regions in the primordial density field.}
{We check that the theoretical large-mass cutoff agrees with the mass functions
measured in numerical simulations. For halos defined by a nonlinear threshold
$\delta=200$ this corresponds to using a linear threshold $\delta_L\simeq 1.59$
instead of the traditional value $\simeq 1.686$.
We also provide a fitting formula that
matches simulations over all mass scales and obeys the exact large-mass tail.
Next, paying attention to the Lagrangian-Eulerian mapping (i.e. corrections
associated with the motions of halos), we improve the standard analytical
formula for the bias of massive halos. We check that our prediction, which
contains no free parameter, agrees reasonably well with numerical simulations.
In particular, it recovers the steepening of the dependence on scale of the bias
that is observed at higher redshifts, which published fitting formulae did not
capture. This behavior mostly arises from nonlinear biasing.}
{}

\keywords{Cosmology: large-scale structure of Universe; gravitation; Methods: analytical}

\maketitle

\section{Introduction} 
\label{Introduction}

The distribution of nonlinear virialized objects, such as galaxies or clusters
of galaxies, is a fundamental test of cosmological models. First, this allows us
to check the validity of the standard cosmological scenario for the formation
of large-scale structures, where nonlinear objects form thanks to the 
amplification by gravitational instability of small primordial density
fluctuations, built for instance by an early inflationary stage 
(e.g., Peebles 1993; Peacock 1998).
For cold dark matter (CDM) scenarios (Peebles 1982), where the amplitude
of the initial perturbations grows at smaller scales, this gives rise to
a hierarchical process, where increasingly large and massive objects form
as time goes on, as increasingly large scales turn nonlinear. 
This process has been largely confirmed by observations, which find smaller
galaxies at very high redshifts (e.g., Trujillo et al. 2007)
while massive clusters of galaxies
(which are the largest bound objects in the Universe) appear at low redshifts
(e.g., Borgani et al. 2001).
Second, on a more quantitative level, statistical properties, such as the
mass function and the two-point correlation of these objects, provide
strong constraints on the cosmological parameters (e.g. through the
linear growth factor $D_+(t)$ of density perturbations) and on the primordial
fluctuations (e.g. through the initial density power spectrum $P_L(k)$).
For these purposes, the most reliable constraints come from observations of
the most massive objects (rare-event tails) at the largest scales.
Indeed, in this regime the formation of large-scale structures is dominated by
the gravitational dynamics (baryonic physics, which involves intricate processes
associated with pressure effects, cooling and heating, mostly occurs at galactic
scales and below), which further simplifies as one probes quasi-linear scales
or rare events where effects associated with multiple mergings can be neglected. 
Moreover, in this regime astrophysical objects, such as galaxies or clusters of
galaxies, can be directly related to dark matter halos, and their abundance
is highly sensitive to cosmological parameters thanks to the steep decline of
the high-mass tail of the mass function (e.g., Evrard 1989).

Thus, the computation of the halo mass function (and especially its large-mass
tail) has been the focus of many works, as it is one of the main properties
measured in galaxy and cluster surveys that can be compared with theoretical
predictions. Most analytical derivations follow the Press-Schechter approach
(Press \& Schechter 1974; Blanchard et al. 1992) or its main extension,
the excursion set theory
of Bond et al. (1991). In this framework, one attempts to estimate the
number of virialized objects of mass $M$ from the probability to have a linear 
density contrast $\delta_L$ at scale $M$ above some given threshold $\delta_c$.
Thus, one identifies current nonlinear halos from positive density fluctuations
in the initial (linear) density field, on a one-to-one basis.
This is rather well justified for rare massive objects, where one can
expect such a link to be valid since such halos should have remained well-defined
objects until now (as they should have suffered only minor mergers).
By contrast, small and typical objects have experienced many mergers and
should be sensitive to highly non-local effects (e.g. tidal forces, mergers),
so that such a direct link should no longer hold, as can be checked in
numerical simulations (Bond et al. 1991).
As noticed by Press \& Schechter (1974), the simplest procedure only yields
half the mass of the Universe in such objects (essentially because only half
of the Gaussian initial fluctuations have a positive density contrast, whatever
the smoothing scale), which they corrected by an ad-hoc multiplicative factor 2.
In this respect the main result of the excursion set theory was to
provide an analytical derivation of this missing factor 2, in the simplified
case of a top-hat filter in Fourier space (Bond et al. 1991). 
Then, it arises from the fact
that objects of mass larger than $M$ are associated with configurations
such that the linear density contrast goes above the threshold $\delta_c$
at some scale $M'\geq M$, which includes cases missed by the Press-Schechter
prescription where the linear density contrast decreases below this
scale $M'$ so that $\delta_L<\delta_c$ at scale $M$ (``cloud-in-cloud'' problem).
The characteristic threshold $\delta_c$ is usually taken as the linear
density contrast reached when the spherical collapse dynamics predicts collapse
to a zero radius. In an Einstein-de Sitter universe this corresponds to
$\delta_c \simeq 1.686$ and to a nonlinear density contrast $\delta \simeq 177$
(assuming full virialization in half the turn-around radius). 
This linear threshold only shows a very weak dependence on cosmological parameters.

Numerical simulations have shown that the Press-Schechter mass function
(PS) is reasonably accurate, especially in view of its simplicity.
Thus, it correctly predicts the typical mass scale of virialized halos
at any redshift, as could be expected since the large-mass behavior is
rather well justified. In addition, it predicts a universal scaling that appears
to be satisfied by the mass functions measured in simulations, that is,
the dependence on halo mass, redshift and cosmology is fully contained
in the ratio $\nu=\delta_c/\sigma(M,z)$, where $\sigma(M,z)$ is the rms linear
density fluctuation at scale $M$. However, as compared with numerical results
it overestimates the low-mass tail and it underestimates the high-mass tail.
This has led to many numerical studies which have provided various fitting
formulae for the mass function of virialized halos, written in terms of
the scaling variable $\nu$ (or $\sigma$) 
(Sheth \& Tormen 1999; Jenkins et al. 2001; Reed et al. 2003; 
Warren et al. 2006; Tinker et al. 2008).
We may note here that a theoretical model that attempts to improve over the PS mass
function is to consider the ellipsoidal collapse dynamics within the excursion-set approach,
to take into account the deviations from spherical symmetry for intermediate mass
halos (Sheth et al. 2001). Note that, as described in the original paper (and emphasized in
Robertson et al. 2009), at large mass this would recover the spherical collapse.
However, the halo mass obtained by such methods is generally underestimated
for non center-of-mass particles (since analytical computations assume that particles
are located at the center of their halo, i.e. they only consider the linear densities within spherical
cells centered on the point of interest). To correct for this effect, in practice one treats
the threshold $\delta_c$ as a free parameter, close to $1.6$, to 
build fitting formulae from numerical simulations. For instance, this correction is contained
in the parameter $a$ in the exponential cutoff of Sheth et al. (2001).

A second property of virialized halos that can be used to constrain 
cosmological models, beyond their number density, is their two-point
correlation. Indeed, observations show that galaxies and clusters do not
obey a Poisson distribution but show significant large-scale correlations
(e.g., McCracken et al. 2008; Padilla et al. 2004).
In particular, their two-point correlation roughly follows the underlying
matter correlation, up to a multiplicative factor $b^2$, called the bias,
that grows for more massive and extreme objects. Following the spirit of
the Press-Schechter picture, Kaiser (1984) found that this behavior arises
in a natural fashion if halos are associated with large overdensities in the
Gaussian initial (linear) density field, above the threshold $\delta_c$.
This was further expanded by Bardeen et al. (1986) and Bond \& Myers (1996),
who considered the clustering of peaks in the Gaussian linear density field.
A simpler derivation, based on a peak-background split argument, and
taking care of the mapping from Lagrangian to Eulerian space, was
presented in Mo \& White (1996). It provides a prediction for the bias
$b(M)$ as a function of halo mass, in the limit of large distance, 
$r\rightarrow \infty$, that agrees reasonably well with numerical simulations.
However, as for the PS mass function, in order to improve the agreement
with numerical results various fitting formulae have been proposed
(Sheth \& Tormen 1999; Hamana et al. 2001; Pillepich et al. 2009).
Again, since the ellipsoidal collapse model reduces to the spherical dynamics
for rare massive halos, it also requires free parameters to improve its accuracy,
but the latter are consistent with those used for the mass function
(Sheth et al. 2001).

In this article we revisit the derivation of the mass function and the bias
of rare massive halos, following the spirit of the Press \& Schechter (1974)
and Kaiser (1984) approaches. That is, we use the fact that large halos
can be identified from overdensities in the Gaussian initial (linear) density
field. First, we briefly review in section~\ref{Spherical-dynamics} some
properties of the growth of linear fluctuations and of the spherical dynamics
in $\Lambda$CDM cosmologies. Next, we recall 
in section~\ref{Density-distribution} that in the quasi-linear regime 
(i.e. at large scales), the probability distribution
of the nonlinear density contrast $\delta_r$ within spherical cells of radius
$r$ can be obtained from spherical saddle-points of a specific action $\cS$,
for moderate values of $\delta_r$ where shell-crossing does not come into play.
We also discuss the properties of these saddle-points as a function of mass,
scale and redshift. Then, we point out in section~\ref{Mass-function} that
this provides the exact exponential tail of the halo mass function.
This applies to any nonlinear density contrast threshold $\delta$ that is used
to define halos, provided it is below the upper bound $\delta_+$ where
shell-crossing comes into play. We compare our results with numerical simulations
and we give a  fitting formula that applies over all mass scales and satisfies
the exact large-mass cutoff. Next, we recall in section~\ref{Halo-profile}
that these results also provide the density profile of dark matter halos
at outer radii (i.e. beyond the virial radius) in the limit of large mass.
Finally, we study the bias of massive halos in section~\ref{Halo-bias},
paying attention to some details such as the Lagrangian-Eulerian mapping, and
we compare our results with numerical simulations. We conclude in 
section~\ref{Conclusion}.

\section{Linear perturbations and spherical dynamics}
\label{Spherical-dynamics}

We consider in this article a flat CDM cosmology with two components,
(i) a non-relativistic component (dark and baryonic matter, which we
do not distinguish here) that clusters through gravitational instability,
and (ii) an uniform dark energy component that does not cluster at the
scales of interest, with an equation-of-state parameter 
$w=p_{\rm de}/\rho_{\rm de}$. 
For the numerical computations we shall focus on a
$\Lambda$CDM cosmology, where the dark energy is associated with a cosmological
constant that is exactly uniform with $w=-1$. However, our results directly 
extend to curved universes (i.e. $\Omega_{\rm k} \neq 0$) and to 
dark energy models with a possibly time-varying $w(z)$, as long as we can
neglect the dark energy fluctuations on the scales of interest, which is valid
for realistic cases.
Focussing on the case of constant $w$, we first recall in this section the
equations that describe the dynamics of the background and of linear matter
density perturbations, as well as the nonlinear spherical dynamics. 

The evolution of the scale factor $a(t)$ is determined
by the Friedmann equation (Wang \& Steinhardt 1998),
\beq
\frac{H^2(t)}{H_0^2} = \Om \, a^{-3} + \Ode \, a^{-3-3w} , \;\; \mbox{with} \;\;
H(t)= \frac{\dot{a}}{a} ,
\label{Friedmann}
\eeq
where subscripts $0$ denote current values at $z=0$, when $a=1$, and a dot
denotes the derivative with respect to cosmic time $t$. 
On the other hand, the density parameters vary with time as
\beq
\Omz(a)= \frac{\Om}{\Om+\Ode \, a^{-3w}} , \;\;
\Odez(a)= \frac{\Ode}{\Om \, a^{3w} +\Ode} .
\label{Omz}
\eeq
Next, introducing the matter density contrast, 
$\delta(\bx,t)=(\rhom(\bx,t)-\rhobm)/\rhobm$,
where $\rhobm(t)$ is the mean matter density, linear density fluctuations 
grow as
\beq
\ddot{\delta}_L + 2 \frac{\dot{a}}{a} \dot{\delta}_L 
- 4\pi\cG\rhobm \delta_L = 0 ,
\label{deltaLt}
\eeq
where the subscript $L$ denotes linear quantities.
For numerical purposes it is convenient to use the logarithm of the scale 
factor as the time variable. Then, using the Friedmann equation (\ref{Friedmann})
the linear growth factor $D_+(t)$ evolves from Eq.(\ref{deltaLt}) as
\beq
D_+'' + \left[ \frac{1}{2}-\frac{3}{2} w \, \Odez \right] D_+' 
- \frac{3}{2} \Omz \, D_+ = 0 ,
\label{D+lna}
\eeq
where we note with a prime the derivative with respect to $\ln a$, as
$D_+'=\dd D_+/\dd\ln a$.
Then, the normalized linear growth factor, $g(t)$, defined as
\beq
g(t) = \frac{D_+(t)}{a(t)}  \;\;\; \mbox{and} \;\;\; g(t=0) =1 ,
\label{gdef}
\eeq
obeys
\beq
g'' + \left[ \frac{5}{2}-\frac{3}{2} w \, \Odez \right] g'
+ \frac{3}{2} (1-w) \Odez \, g = 0 ,
\label{glna}
\eeq
with the initial conditions $g\rightarrow 1$ and  $g' \rightarrow 0$
at $a\rightarrow 0$. 

In the following we shall also need the dynamics of spherical density
fluctuations. For such spherically symmetric initial conditions, the physical
radius $r(t)$, that contains the constant mass $M$ until shell-crossing, 
evolves as
\beq
\ddot{r} = -\frac{4\pi\cG}{3} \, r \, [ \rhom+(1+3w)\rhobde ] \;\;\; 
\mbox{with} \;\;\; \rhom = \frac{3M}{4\pi r^3} .
\label{rt}
\eeq
Note that $\rhom$ is the mean density within the sphere of radius $r$
(and not the local density at radius $r$).
Using again the Friedmann equation (\ref{Friedmann}) this reads as
\beq
r'' - \frac{3}{2} (1+w \, \Odez) r' 
+ \frac{\rhom+(1+3w)\rhobde}{2(\rhobm+\rhobde)} \, r = 0 .
\label{rtau}
\eeq
As for the linear growth factor $D_+(t)$, it is convenient to introduce the
normalized radius $y(t)$ defined as
\beq
y(t) = \frac{r(t)}{q(t)} \;\;\; \mbox{and} \;\;\; q(t) \propto a(t), 
\;\;\;y(t=0) = 1 .
\label{ydef}
\eeq
Thus, $q(t)$ is the Lagrangian coordinate of the shell $r(t)$, that is, the
physical radius that would enclose the same mass $M$ in a uniform universe
with the same cosmology. This also implies that the density $\rhom$ writes as
\beq
\rhom(t) = \rhobm(t) \, y(t)^{-3} .
\label{rhomy}
\eeq
Then, Equation (\ref{rtau}) leads to
\beq
y'' + \left[\frac{1}{2} - \frac{3}{2} w \, \Odez \right] y'
+ \frac{\Omz}{2} \left(y^{-3}-1\right) y = 0 .
\label{ylna}
\eeq
Of course, we can check that in the linear regime, where $y_L= 1-\delta_L/3$,
we recover Eq.(\ref{D+lna}) for the linear growth of $\delta_L$.
Then, to obtain the nonlinear density contrast, $\delta(z)=\rhom/\rhobm-1$,  
associated with the linear density contrast, $\delta_L(z)$, at a given redshift
$z$, we solve Eq.(\ref{ylna}) with the initial condition $y(\zi)=1-\deltaLi/3$ 
and $y'(\zi)= -\deltaLi/3$ at some high redshift $\zi$,
with $\deltaLi/\delta_L=D_+(\zi)/D_+(z)$.
For any redshift $z$ this defines a mapping
$\delta_L\mapsto \delta=\cF(\delta_L)$ that fully describes the spherical 
dynamics before shell-crossing.

\begin{figure}[htb]
\begin{center}
\epsfxsize=8 cm \epsfysize=5.5 cm {\epsfbox{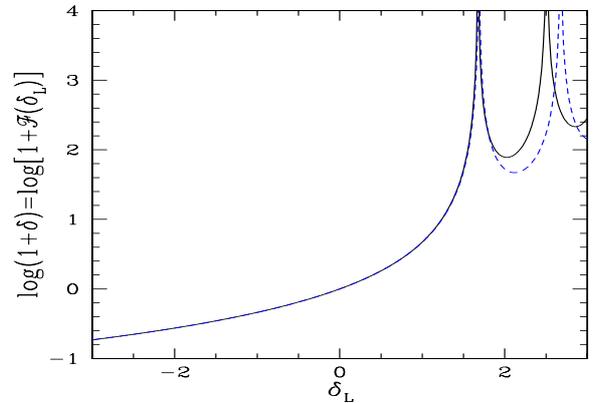}}
\end{center}
\caption{The function $\cF(\delta_L)$ that describes the spherical dynamics
when there is no shell-crossing, at $z=0$. The solid line corresponds to the
$\Lambda$CDM cosmology (with $\Om=0.27$) and the dashed line to the
Einstein-de Sitter case $\Om=1$. The second peak corresponds to a second 
collapse to the center, but for our purposes we only need $\cF(\delta_L)$
before first collapse.}
\label{figlrho_deltaL}
\end{figure}

\begin{figure}[htb]
\begin{center}
\epsfxsize=8 cm \epsfysize=6. cm {\epsfbox{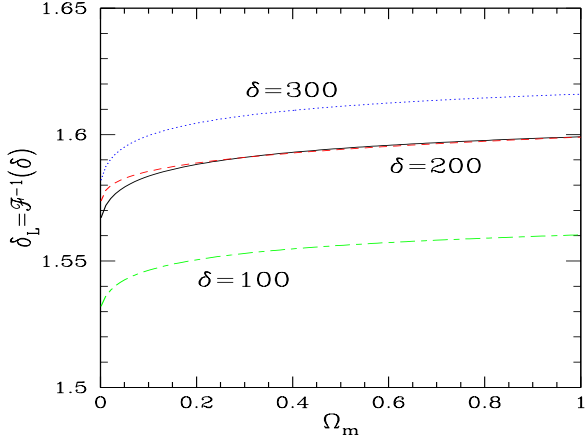}}
\end{center}
\caption{The linear density contrast $\delta_L=\cF^{-1}(\delta)$ associated
with the nonlinear density contrast $\delta$, through the spherical
dynamics, as a function of the cosmological parameter $\Omz(z)$ at the
redshift of interest. We show the three cases $\delta=100,200$ and $300$.
The dashed line is the fit (\ref{deltaL200fit}) to the case $\delta=200$.}
\label{figdeltaLOm}
\end{figure}

We compare in Fig.~\ref{figlrho_deltaL} the function $\cF(\delta_L)$ obtained
at $z=0$ within the $\Lambda$CDM cosmology that we consider in this article (solid
line) with the Einstein-de Sitter case (dashed line), where it has a well-known
parametric form (Peebles 1980). As is well known, we can check that the
dependence on $\Om$ is very weak until full collapse to a point, which occurs
at slightly lower values of $\delta_L$ for low $\Om$.
We show in Fig.~\ref{figdeltaLOm}
the linear density contrasts, $\delta_L=\cF^{-1}(\delta)$, associated with
three nonlinear density contrasts, $\delta=100,200$ and $300$, as a function
of the cosmological parameter $\Omz(z)$. In agreement with 
Fig.~\ref{figlrho_deltaL}, they show a slight decrease for smaller $\Omz$.
For $\delta=200$ a simple fit (dashed line) is provided by
\beq
\delta=200: \;\; \delta_L \simeq 1.567 + 0.032 \, (\Omz+0.0005)^{0.24} ,
\label{deltaL200fit}
\eeq
which agrees with the exact curve to better than $5\times 10^{-4}$ for
$\Omz>0.2$, which covers the range of practical interest.

In this article we consider Gaussian initial fluctuations, which are fully
defined by the linear density power spectrum
\beq
\lag \tdelta_L(\bk_1)\tdelta_L(\bk_2) \rag = \delta_D(\bk_1+\bk_2) \, P_L(k_1) ,
\label{PLdef}
\eeq
where $\delta_D$ is the Dirac distribution and we normalized the Fourier
transform as
\beq
\delta(\bx) = \int\dd\bk \, e^{\ii\bk.\bx} \, \tdelta(\bk) ,
\label{Fourierdef}
\eeq
where $\bx$ and $\bk$ are the comoving spatial coordinate and wavenumber.
This gives for the linear density two-point correlation
\beqa
C_L(\bx_1,\bx_2) & = & \lag \delta_L(\bx_1) \delta_L(\bx_2) \rag \nonumber \\
& = & 4\pi \int \dd k \, k^2 \, P_L(k) \, 
\frac{\sin(k|\bx_2-\bx_1|)}{k|\bx_2-\bx_1|} .
\label{CLdef}
\eeqa

We also introduce the smoothed density contrast, $\delta_r(\bx)$, within the
sphere of radius $r$ and volume $V$ around position $\bx$,
\beq
\delta_r(\bx) = \int_V \frac{\dd \bx'}{V} \, \delta(\bx+\bx') 
= \int\dd\bk \, e^{\ii\bk.\bx} \, \tdelta(\bk) \, \tW(kr) ,
\label{deltardef}
\eeq
with a top-hat window that reads in Fourier space as 
\beq
\tW(kr) = \int_V \frac{\dd \bx}{V} \, e^{\ii\bk.\bx} = 
3 \frac{\sin(kr)-kr\cos(kr)}{(kr)^3} .
\label{Wdef}
\eeq
Then, in the linear regime, the cross-correlation of the smoothed linear density
contrast at scales $r_1$ and $r_2$ and positions $\bx_1$ and $\bx_2=\bx_1+\bx$
reads as
\beqa
\sigma^2_{r_1,r_2}(x) & = & \lag \delta_{Lr_1}(\bx_1)
\delta_{Lr_2}(\bx_1+\bx) \rag  \nonumber \\
& = & 4\pi\int\dd k \, k^2 P_L(k) \tW(kr_1) \tW(kr_2) 
\frac{\sin(kx)}{kx} .
\label{sigr1r2}
\eeqa
In particular, $\sigma_r=\sigma_{r,r}(0)$ is the usual rms linear density
contrast at scale $r$.

\section{Distribution of the density contrast}
\label{Density-distribution}

We recall here that in the quasi-linear limit, $\sigma_r\rightarrow 0$,
the distribution $\cP(\delta_r)$ of the nonlinear density contrast $\delta_r$
at scale $r$ can be derived from a steepest-descent method (Valageas 2002a).
In agreement with the alternative approach of Bernardeau (1994a),
this shows that rare-event tails are dominated by spherical saddle-points,
which we use in sections~\ref{Mass-function}-\ref{Halo-bias} to obtain the
properties of massive halos.

Since the system is statistically homogeneous we can consider the sphere of
radius $r$ centered on the origin $\bx=0$.
Then, we first introduce the cumulant generating function $\varphi(y)$,
\beq
e^{-\varphi(y)/\sigma_r^2} = \lag e^{-y\delta_r/\sigma_r^2} \rag 
= \int_{-1}^{\infty} \dd\delta_r \, e^{-y\delta_r/\sigma_r^2} \,
\cP(\delta_r) ,
\label{phidef}
\eeq
which determines the distribution $\cP(\delta_r)$ through the inverse
Laplace transform
\beq
\cP(\delta_r) = \inta \frac{\dd y}{2\pi\ii\sigma_r^2} \, 
e^{[y\delta_r-\varphi(y)]/\sigma_r^2} .
\label{Pphi}
\eeq
In Eq.(\ref{phidef}) we rescaled the cumulant generating function by a
factor $\sigma_r^2$ so that it has a finite limit in the quasi-linear limit
$\sigma_r\rightarrow 0$ for the Gaussian initial density fluctuations that
we study in this article (Bernardeau et al. 2002). 
In particular, its expansion at $y=0$ reads as
\beq
\varphi(y) = - \sum_{p=2}^{\infty} \frac{(-y)^p}{p!} \, 
\frac{\lag\delta_r^p\rag_c}{\sigma_r^{2(p-1)}} .
\label{phiy0}
\eeq
Then, the average (\ref{phidef}) can be written as the path integral
\beq
e^{-\varphi(y)/\sigma_r^2} = (\det C_L^{-1})^{1/2} \int\cD \delta_L(\bq)
\; e^{-\cS[\delta_L]/\sigma_r^2} ,
\label{path}
\eeq
where $C_L^{-1}$ is the inverse matrix of the two-point correlation 
(\ref{CLdef}) and the action $\cS$ reads as
\beq
\cS[\delta_L] = y \, \delta_r[\delta_L] + \frac{\sigma_r^2}{2} \,
\delta_L . C_L^{-1} . \delta_L
\label{Sdef}
\eeq
Here $\delta_r[\delta_L]$ is the nonlinear functional that affects to the
initial condition defined by the linear density field $\delta_L(\bq)$ the
nonlinear density contrast $\delta_r$ within the sphere of radius $r$, and
we note $\delta_L . C_L^{-1} . \delta_L = \int \dd\bq_1\dd\bq_2 
\delta_L(\bq_1)C_L^{-1}(\bq_1,\bq_2)\delta_L(\bq_2)$. 
Note that the action $\cS$ does not depend on the normalization of the linear
power spectrum since both $\sigma_r^2$ and $C_L$ are proportional to
$P_L$.

\begin{figure}[htb]
\begin{center}
\epsfxsize=8.6 cm \epsfysize=6. cm {\epsfbox{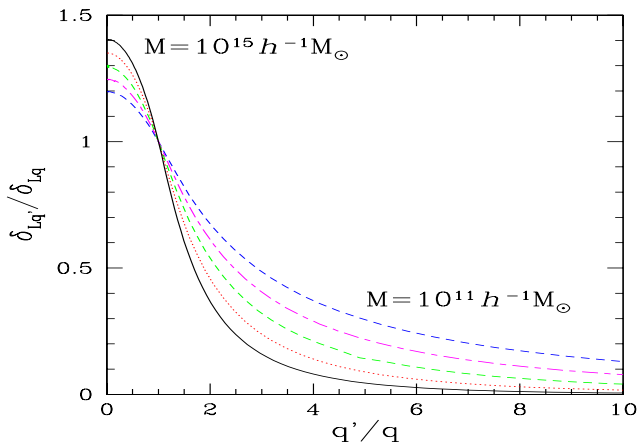}}
\end{center}
\caption{The radial profile (\ref{profile}) of the linear density contrast
$\delta_{Lq'}$ of the saddle point of the action $\cS[\delta_L(\bq)]$.
We show the profiles obtained with a $\Lambda$CDM cosmology for the
masses $M=10^{11},10^{12},10^{13},10^{14}$ and $10^{15}h^{-1}M_{\odot}$.
A larger mass corresponds to a lower ratio $\delta_{Lq'}/\delta_{Lq}$
at large radii $q'/q>1$.}
\label{figdeltaLq}
\end{figure}

In the quasi-linear limit, $\sigma_r\rightarrow 0$, as shown in Valageas (2002a)
the path integral (\ref{path}) is dominated by the minimum\footnote{As described
in details in Valageas (2002a), depending on the slope of the linear power
spectrum the saddle-point (\ref{profile}) may only be a local minimum or 
maximum, but it still governs the rare-event limit, in agreement with physical
expectations.} of the action $\cS[\delta_L]$,
\beq
\sigma_r \rightarrow 0 : \;\; \varphi(y) \rightarrow \min_{\delta_L(\bq)}
\cS[\delta_L] .
\label{Smin}
\eeq
Using the spherical symmetry of the top-hat window $W$, in agreement with
the approach of Bernardeau (1994a), one obtains a spherical saddle-point
with the radial profile
\beq
\delta_{Lq'} = \delta_{Lq} \, \frac{\sigma^2_{q,q'}}{\sigma^2_q} ,
\label{profile}
\eeq
where $\sigma_{q,q'}=\sigma_{q,q'}(0,0)$ and $q$ is the Lagrangian
coordinate associated with the Eulerian radius $r$ through the spherical
dynamics recalled in section~\ref{Spherical-dynamics}, and $q'$ is a dummy
Lagrangian coordinate along the radial profile (hereafter we denote
by the letter $q$ Lagrangian radii, which are associated with the
linear density field, whereas $r$ denotes Eulerian radii associated with
the nonlinear density contrast $\delta_r$). Thus, the radii $q'$ and $r'$ are
related by
\beq
q'^3 = (1+\delta_{r'}) \, r'^3 , \;\;\; \mbox{with} \;\;\;
\delta_{r'} = \cF(\delta_{Lq'}) ,
\label{qr}
\eeq
where the function $\delta_{r'}=\cF(\delta_{Lq'})$, that describes the
spherical dynamics, was obtained below Eq.(\ref{ylna}) and shown in 
Fig.~\ref{figlrho_deltaL}.

We can note that the profile (\ref{profile}) is also the mean conditional
profile of the linear density contast $\delta_{Lq'}$, under the constraint
that is it equal to a given value $\delta_{Lq}$ at a given radius $q$ 
(e.g., Bernardeau 1994a). The reason why the nonlinear dynamics gives back
this result is that the nonlinear density contrast $\delta_r$ only depends
on the linear density contrast $\delta_{Lq}$ at the associated Lagrangian
radius $q$, through the mapping $\delta_r=\cF(\delta_{Lq})$. Indeed, as long
as shell-crossing does not modify the mass enclosed within the shell of
Lagrangian coordinate $q$, its dynamics is independent of the motion of
inner and outer shells (thanks to Gauss theorem). Then, in order to obtain
the minimum of the action $\cS$ we could proceed in two steps. First,
for arbitrary Lagrangian radius $q$ and linear contrast $\delta_{Lq}$,
we minimize $\cS$ with respect to the profile $\delta_{Lq'}$ over $q'\neq q$.
From the previous argument, only the second Gaussian term in (\ref{Sdef})
varies so that this partial minimization leads to the profile (\ref{profile})
(indeed, for Gaussian integrals the saddle-point method is exact).
Second, we minimize over the Lagrangian radius $q$ (or equivalently over
$\delta_L$ or $\delta_r$), which leads to Eq.(\ref{phiminG}) below.
Here we also use the fact that a spherical saddle-point with respect to
spherical fluctuations is automatically a saddle-point with respect to
arbitrary non-spherical perturbations and it can be seen that for small
$y$ one obtains a minimum (then we assume that at finite $y$
strong deviations from spherical symmetry do not give rise to deeper minima,
which seems natural from physical expectations).
We refer to Valageas (2002a,2009b) for more detailed derivations.

Note that the shape of the linear
profile (\ref{profile}) depends on the shape of the linear power spectrum,
whence on the mass scale $M$ of the saddle point for a curved CDM
linear power spectrum, but not on redshift.
We show in Fig.~\ref{figdeltaLq} the profile (\ref{profile}) obtained for
several masses $M$. For a power-law linear power spectrum, of slope $n$,
Eq.(\ref{profile}) leads to $\delta_{Lq'} \propto q'^{-(n+3)}$ at large
radii, $q'\rightarrow\infty$. Then, since for a CDM cosmology $n$ increases
at larger scales, the profile shows a steeper falloff at large radii for
larger mass, in agreement with Fig.~\ref{figdeltaLq} (in this section we
consider a $\Lambda$CDM cosmology with 
$(\Om,\OL,\sigma_8,n_s,h)=(0.27,0.73,0.79,0.95,0.7)$).

\begin{figure}[htb]
\begin{center}
\epsfxsize=8.6 cm \epsfysize=6. cm {\epsfbox{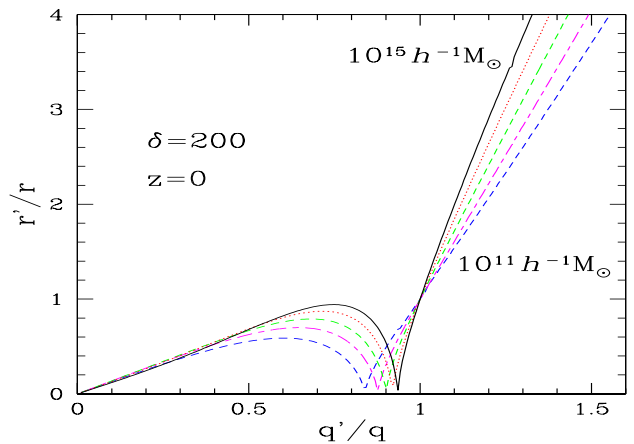}}
\end{center}
\caption{The Lagrangian map $q'\mapsto r'$ given by the spherical dynamics
(neglecting shell-crossing) for the saddle-point (\ref{profile}) with
a nonlinear density contrast $\delta=200$ at the Eulerian radius $r$.
We plot the curves obtained at $z=0$ for several masses, as in 
Fig.~\ref{figdeltaLq}. Inner shells have already gone once through the
center (hence their dynamics is no longer exactly given by Eq.(\ref{ylna}))
but they have not crossed the radius $r$ yet.}
\label{figrq}
\end{figure}

We show in Fig.~\ref{figrq} the Lagrangian map, $q'\mapsto r'$,
given by the spherical dynamics (i.e. the function $\cF(\delta_L)$ where
we neglect shell-crossing) for the saddle-point (\ref{profile}), with
a nonlinear density contrast $\delta=200$ at the Eulerian radius $r$,
at redshift $z=0$. 
Inner shells have already gone once through the center but they have not
reached radius $r$ yet. Even though their dynamics is no longer exactly given
by Eq.(\ref{ylna}), an exact computation would give the same property as
the increasing mass seen by these particles, as they pass outer shells, should
slow them down as compared with the constant-mass dynamics.
In agreement with Fig.~\ref{figdeltaLq}, for larger masses, which have a larger
central linear density contrast, shell-crossing has moved to larger radii
(the local maximum of $r'/r$, to the left of $r'=0$, is higher).

\begin{figure}[htb]
\begin{center}
\epsfxsize=8.6 cm \epsfysize=6. cm {\epsfbox{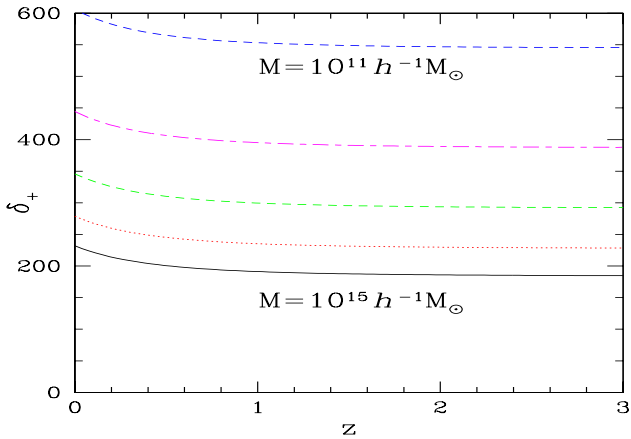}}
\end{center}
\caption{The nonlinear density contrast $\delta_+$, beyond which shell-crossing
must be taken into account, as a function of redshift for several mass scales.}
\label{figshellcrossz}
\end{figure}

From the Lagrangian map, $q'\mapsto r'$, we define the nonlinear density
threshold, $\delta_+$, as the nonlinear density contrast $\delta_r$ reached
within radius $r$ at the time when inner shells first cross this radius $r$.
Then, up to $\delta_+$, the mass within the Lagrangian shell $q$ has remained
constant, so that the saddle-point (\ref{profile})-(\ref{qr}) is exact
(this slightly underestimates $\delta_+$ as the expansion of inner shells
should be somewhat slowed down by the mass of the outer shells they have
overtaken). We show in Fig.~\ref{figshellcrossz} the dependence on redshift
of this threshold $\delta_+$, for several masses. In agreement with 
Figs.~\ref{figdeltaLq}, \ref{figrq}, this threshold is smaller for larger
masses.
It shows a slight decrease at higher redshift as $\Omz(z)$ grows to unity.
We can see that for massive clusters at $z=0$, which have a mass of order
$10^{15}h^{-1}M_{\odot}$, the density contrast $\delta\simeq 200$ should
separate outer shells with radial accretion from inner shells with a significant
transverse velocity dispersion, built by the radial-orbit instability that
dominates the dynamics after shell-crossing, see Valageas 2002b (in particular
the appendix A). This agrees with numerical simulations, see for instance the
lower panel of Fig.3 in Cuesta et al. (2008), which show that rare massive
clusters exhibit a strong radial infall pattern, with a low velocity dispersion,
beyond the virial radius (where $\delta_r\sim 200$), while inner radii show
a large velocity dispersion (even though we can distinguish close to the virial
radius the outward velocity associated with the shells that have gone once
through the center).
Small-mass halos do not show such a clear infall pattern and the velocity
dispersion is significant at all radii (e.g., upper panel of Fig.3 in
Cuesta et al. 2008). This expresses the fact that such halos, associated with
typical events and moderate density fluctuations, are no longer governed
by the spherical saddle-point (\ref{profile}). Indeed, at low mass and small
Lagrangian radius $q$, $\sigma_q$ is no longer very small so that the
path integral (\ref{path}) is no longer dominated by its minimum and one
must integrate over all typical initial conditions, including large deviations
from spherical symmetry.

Next, the amplitude $\delta_{Lq}$ and the minimum $\varphi$
are given as functions of $y$ by the implicit equations (Legendre transform)
\beq
y= -\tau \frac{\dd\tau}{\dd\cG}  \;\;\; \mbox{and} \;\;\;
\varphi= y \cG + \frac{\tau^2}{2} ,
\label{yphi}
\eeq
where the function $\tau(\cG)$ is defined from the spherical dynamics
through the parametric system (Valageas 2002a; Bernardeau 1994b)
\beq
\cG=\delta_r=\cF(\delta_{Lq})  \;\;\;  \mbox{and} \;\;\;
\tau=-\delta_{Lq} \frac{\sigma_r}{\sigma_q} .
\label{Gtau}
\eeq
The system (\ref{yphi}) also reads as
\beq
\varphi=\min_{\cG}\left[ y \cG + \frac{\tau^2(\cG)}{2}\right] .
\label{phiminG}
\eeq
Note that the function $\tau(\cG)$, whence the cumulant generating function
$\varphi(y)$, depends on the shape of the linear power spectrum through the
ratio of linear power $\sigma_r/\sigma_q$ at Eulerian and Lagrangian
scales $r$ and $q$.
Finally, from the inverse Laplace transform (\ref{Pphi}), a second
steepest-descent integration over the variable $y$ gives (Valageas 2002a)
\beq
\sigma_r \rightarrow 0 : \;\; \cP(\delta_r) \sim e^{-\tau^2/(2\sigma_r^2)} 
= e^{-\delta_{Lq}^2/(2\sigma_q^2)} .
\label{PdL}
\eeq
Thus, in the quasi-linear limit, the distribution $\cP(\delta_r)$ is governed
at leading order by the Gaussian weight of the linear fluctuation $\delta_{Lq}$
that is associated with the nonlinear density contrast $\delta_r$ through the
spherical dynamics. We can note that Eq.(\ref{PdL}) could also be obtained
from a Lagrange multiplier method, without introducing the generating function
$\varphi(y)$. Indeed, in the rare-event limit, where $\cP(\delta)$ is governed
by a single (or a few) initial configuration, we may write
\beq
\mbox{rare events}: \;\;  \cP(\delta) \sim \max_{\{\delta_L[\bq]{\displaystyle |}
\delta_r[\delta_L]=\delta\}} e^{-\frac{1}{2} \delta_L.C_L^{-1}.\delta_L} .
\label{rareP}
\eeq
That is, $\cP(\delta)$ is governed by the maximum of the Gaussian weight
$e^{-(\delta_L.C_L^{-1}.\delta_L)/2}$ subject to the constraint 
$\delta_r[\delta_L]=\delta$ (assuming there are no degenerate maxima).
Then, we can obtain this maximum by minimizing the action 
$\cS[\delta_L]/\sigma_r^2$ of Eq.(\ref{Sdef}), where $y$ plays the role of
a Lagrange multiplier. This gives the saddle-point (\ref{profile}), and
the amplitude $\delta_{Lq}$ and the radius $q$ are directly obtained from the
constraint $\delta=\cF(\delta_{Lq})$, as in Eq.(\ref{qr}). Then, we do not
need the explicit expression of the Lagrange multiplier $y$, as this is sufficient
to obtain the last expression of the asymptotic tail (\ref{PdL}).
Nevertheless, it is useful to introduce the generating function $\varphi(y)$,
which makes it clear that the Lagrange multiplier $y$ is also the Laplace
conjugate of the nonlinear density contrast $\delta$ as in Eq.(\ref{phidef}),
since it is also of interest by itself, as it yields the density cumulants through
the expansion (\ref{phiy0}).

The asymptotic (\ref{PdL}) holds in the rare-event limit.
This corresponds to both the quasi-linear limit, $\sigma_r\rightarrow 0$ at
fixed density contrast $\delta_r$, and to the low-density limit, 
$\delta_r\rightarrow 0$ at fixed $\sigma_r$, as long as there is no
shell-crossing. This latter requirement gives a lower boundary $\delta_-$,
for linear power spectra with a slope $n>-1$, and an upper boundary
$\delta_+$, for any linear spectrum (Valageas 2002b).
This upper boundary was shown in Fig.~\ref{figshellcrossz} for a $\Lambda$CDM
cosmology, for several masses.

Indeed, at large positive density contrast, shell-crossing always occurs,
as seen in Figs.~\ref{figrq}-\ref{figshellcrossz}.
This invalidates the mapping $\delta_L\mapsto\delta=\cF(\delta_L)$ obtained
from Eq.(\ref{ylna}), as mass is no longer conserved within the Lagrangian
shell $q$, so that the asymptotic behavior (\ref{PdL}) is no longer exact.
In fact, as shown in Valageas (2002b), after shell-crossing it is no longer
sufficient to follow the spherical dynamics, even if we take into account
shell-crossing. Indeed, a strong radial-orbit instability develops so that the
sensitivity to initial perturbations actually diverges when particles cross
the center of the halo. Then, the functional $\delta_r[\delta_L(\bq)]$ is
singular at such spherical states (i.e. it is discontinuous as infinitesimal 
deviations from spherical symmetry lead to a finite change of $\delta_r$)
and the path integral (\ref{path}) is no longer governed by spherical states
that have a zero measure. As noticed above, this
also means that, in the limit of rare massive halos, the nonlinear density
threshold $\delta_+$ separates outer shells with a smooth radial flow
from inner shells with a significant transverse velocity dispersion.
Thus, $\delta_+$ marks the virialization radius where isotropization of the
velocity tensor becomes important, in agreement with numerical simulations
(e.g., Cuesta et al. 2008).

It is interesting to note that a similar approach can be developed for
the ``adhesion model'' (Gurbatov et al. 1989), where particles move according
to the Zeldovich dynamics (Zeldovich 1970) but do not cross because of
an infinitesimal viscosity (i.e. they follow the Burgers dynamics in the
inviscid limit). Moreover, in the one-dimensional case, with a linear
power-spectrum slope $n=-2$ or $n=0$ (i.e. the linear velocity is a Brownian
motion or a white noise), it is possible to derive the exact distribution
$\cP(\delta_r)$ by other techniques and to check that it agrees with the
asymptotic tail (\ref{PdL}), as seen in Valageas (2009a,b).

\section{Mass function of collapsed halos}
\label{Mass-function}

The quasi-linear limit (\ref{PdL}) of the distribution $\cP(\delta_r)$
clearly governs the large-mass tail of the mass function $n(M)\dd M$,
where we define halos as spherical objects with a fixed density contrast
$\delta_r=\delta$. Indeed, since the Lagrangian radius $q$ is related to the
halo mass by $M=\rhobm 4\pi q^3/3$, massive halos correspond to large Eulerian
and Lagrangian radii, and the limit $M\rightarrow\infty$ corresponds to the
quasi-linear limit $\sigma_r\rightarrow 0$. Then, going from the distribution
$\cP(\delta_r)$ to the mass function $n(M)$ can introduce geometrical 
power-law prefactors since halos are not exactly centered on the cells of a
fixed grid, as discussed in Betancort-Rijo \& Montero-Dorta (2006), 
but the exponential cutoff remains the same as in Eq.(\ref{PdL}), whence
\beq
M \rightarrow \infty : \;\;\; \ln[n(M)] \sim - \frac{\delta_L^2}{2\sigma^2} ,
\;\;\; \mbox{with} \;\; \delta_L=\cF^{-1}(\delta) ,
\label{nMtail}
\eeq
where $\sigma=\sigma_q$ with $M=\rhobm 4\pi q^3/3$.
A simple approximation for the mass function that satisfies this large-mass
falloff can be obtained from the Press-Schechter method (PS), see
Press \& Schechter (1974), but using the actual
linear threshold $\delta_L=\cF^{-1}(\delta)$ associated with the
nonlinear threshold $\delta$ that defines the halo radius $r$, rather than
the linear threshold $\delta_c \simeq 1.686$ associated with the spherical
collapse down to a point. 
This yields for the number of halos in the range $[M,M+\dd M]$ per unit volume,
\beq
n(M) \, \dd M = \frac{\rhobm}{M} \, f(\sigma) \, 
\left|\frac{\dd \sigma}{\sigma}\right| ,
\label{nM}
\eeq
with
\beq
f(\sigma) = \sqrt{\frac{2}{\pi}} \, \frac{\delta_L}{\sigma} \,
e^{-\delta_L^2/(2\sigma^2)} .
\label{fPSdL}
\eeq
The mass function (\ref{fPSdL}) includes the usual prefactor $2$ that gives
the normalization
\beq
\int_0^{\infty} \frac{\dd\sigma}{\sigma} \, f(\sigma) = 1 ,
\label{normf}
\eeq
which ensures that all the mass is contained in such halos.
However, we must note that the power-law prefactor in (\ref{fPSdL}) has
no strong theoretical justification since only the exponential cutoff
(\ref{nMtail}) was exactly derived from (\ref{PdL}). 
In principle, it could be possible to
derive subleading terms (whence power-law prefactors) in the quasi-linear
limit for $\cP(\delta_r)$, by expanding the path integral (\ref{path})
around the saddle-point (\ref{profile}). Then, one would also need to
take more care of the prefactors that would arise as one goes from the
distribution $\cP(\delta_r)$ to the mass function $n(M)$. However, this
expansion would not allow one to derive the shape of the mass function at
low masses, as this regime is far from the quasi-linear limit and involves
multiple mergers far from spherical symmetry.
Until a new method is devised to handle this regime, one must treat the
prefactors as free parameters to be fitted to numerical simulations, in order
to describe the mass function from small to large masses.

Here we can note that in the Press-Schechter approach (and in most models)
the mass function (\ref{fPSdL}) is obtained from the linear density reached within
the sphere centered on each mass element. That is, a particle is assumed to
be part of a halo of mass larger than $M$ if the sphere of mass $M$ (or a sphere
of mass larger than $M$ in the excursion set approach of Bond et al. 1991),
centered on this particle, has a linear density contrast above a threshold $\delta_L$.
As pointed out
in Sheth, Mo \& Tormen (2001), and discussed in their section~3, since all particles
are not located at the center of their parent halo, the correct criteria should rather
be that the particle belongs to a sphere, not necessarily centered on this point, of
mass greater than $M$, that has collapsed by the redshift of interest. Within the
excursion set approach of Bond et al. (1991), this means that one should consider
the first-crossing distributions associated with center-of-mass particles, rather
than with randomly chosen particles, as argued in Sheth, Mo \& Tormen (2001).
Then, the latter suggest that this could modify the numerical factor in the exponential
tail of the mass function, that is, lead to a smaller factor $\delta_L$ in Eq.(\ref{nMtail})
than the one associated with spherical (or ellipsoidal) collapse dynamics.
We point out here that this effect should only give subleading corrections to the tail
(\ref{nMtail}), so that the factor $\delta_L$ in Eq.(\ref{nMtail}) remains exactly given
by the spherical collapse dynamics.

This can be seen from the fact that the same effect would apply to the density
probability distribution $\cP(\delta_r)$, as randomly placed cells of radius $r$ are
typically not centered on halo profiles. Nevertheless, the results
(\ref{yphi})-(\ref{rareP}) are exact in the quasi-linear limit (as can also be checked
by the comparison with standard perturbation theory for the cumulant generating
function $\varphi(y)$), and off-center effects would be included in the subleading
terms, computed as usual by expanding the path integral (\ref{path}) around its
saddle-point, which would typically give the power-law prefactor to the tail
(\ref{PdL}). In terms of the halo mass function itself, this point was also studied in
Betancort-Rijo \& Montero-Dorta (2006), who found as expected that at large mass
such a geometrical factor only modifies that power-law prefactor.

As for the probability distribution $\cP(\delta_r)$, it is interesting to
note that a similar high-mass tail can be derived for the ``adhesion model'',
where halos are defined as zero-size objects (shocks). Again, in the
one-dimensional case, for both $n=-2$  and $n=0$, where the exact mass function
can be obtained by other means, one can check that it agrees with the
analog of the asymptotic tail (\ref{nMtail}) (Valageas 2009a,b,c).
Thus, we can check that in these two non-trivial examples the leading-order terms
for the large-mass decays of $\cP(\delta_r)$ and $n(m)$ are exactly set by
saddle-point properties and are not modified by the off-center effects discussed
above.

We can note that the explanation of the large-mass tail (\ref{nMtail})
by the exact asymptotic result of the steepest-descent method described in
section~\ref{Density-distribution} agrees nicely with numerical simulations.
This is most clearly seen in Figs.~3 and 6 of Robertson et al. (2009), 
who trace back the linear density contrast $\delta_L$ of the Lagrangian
regions that form halos at $z=0$. Their results show that the distribution
of linear contrasts $\delta_L$, measured as a function of mass 
(or of $\sigma(M)$), has a roughly constant lower bound $\delta_L^-$, with
$\delta_L^-\sim 1.6$, and an upper bound 
$\delta_L^+$that grows with $\sigma(M)$. 
We must note however that the difference between $1.59$ and $1.6754$
(which would be the standard threshold in their $\Lambda$CDM cosmology
with $\Om=0.27$) is too small to discriminate both values from the results
shown in their figures, so that these numerical simulations alone do not
give the asymptotic value $\delta_L^-$ to better than about $0.2$.
They obtain the same results when they define halos
by nonlinear density contrasts $\delta=100$ or $\delta=600$, with a lower
bound $\delta_L^-$ that grows somewhat with $\delta$. In terms of the approach
described above, this behavior expresses the fact that the most probable
way to build a massive halo of nonlinear density contrast $\delta$ is
to start from a Lagrangian region of linear density contrast
$\delta_L=\cF^{-1}(\delta)$, which obeys the spherical profile (\ref{profile})
and corresponds to the saddle-point of the action (\ref{Sdef}). As recalled
in section~\ref{Density-distribution}, the path integral (\ref{path}) is
increasingly sharply peaked around this initial state at large mass scales,
which explains why the dispersion of linear density contrasts $\delta_L$
measured in the simulations decreases with $\sigma(M)$. At smaller mass,
one is sensitive to an increasingly broad region around the saddle-point,
which mostly includes non-spherical initial fluctuations. Since these initial
conditions are less efficient to concentrate matter in a small region
one needs a larger linear density contrast $\delta_L$ to reach the same
nonlinear threshold $\delta$ within the Eulerian radius $r$, which is why
the distribution is not symmetric and mostly broadens by increasing its
typical upper bound $\delta_L^+$. In principles, it may be possible to estimate
the width of this distribution, at large masses, by expanding the action
$\cS[\delta_L]$ around its saddle-point.

\begin{figure}[htb]
\begin{center}
\epsfxsize=9 cm \epsfysize=7 cm {\epsfbox{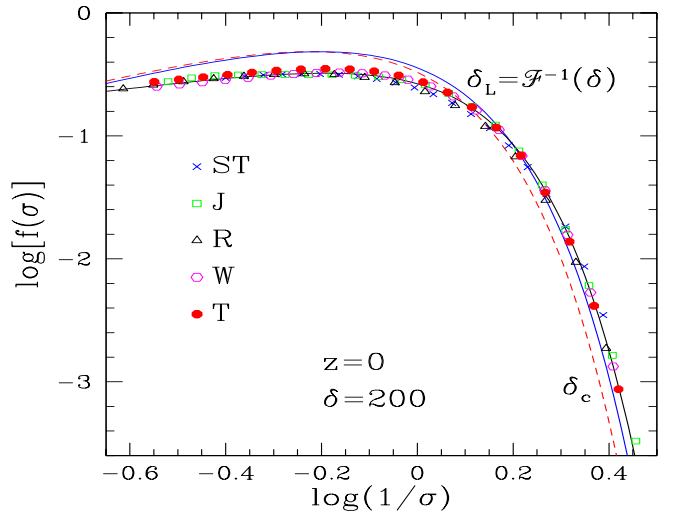}}
\end{center}
\caption{The mass function at redshift $z=0$ of halos defined by the nonlinear
density contrast $\delta=200$. The points are the fits to numerical
simulations from Sheth \& Tormen (1999), Jenkins et al. (2001),
Reed et al. (2003), Warren et al. (2006) and Tinker et al. (2008). 
The dashed line (``$\delta_c$'') is the usual Press-Schechter mass function,
while the solid line that goes close to the Press-Schechter prediction at small
mass is the mass function (\ref{fPSdL}) with the exact cutoff (\ref{nMtail}).
Here we have $\delta_L \simeq 1.59$.
The second solid line that agrees with simulations
over the whole range is the fit (\ref{fitfsig}), that obeys the same large-mass
exponential cutoff.}
\label{figfsig_z0_D200}
\end{figure}

\begin{figure}[htb]
\begin{center}
\epsfxsize=9 cm \epsfysize=6 cm {\epsfbox{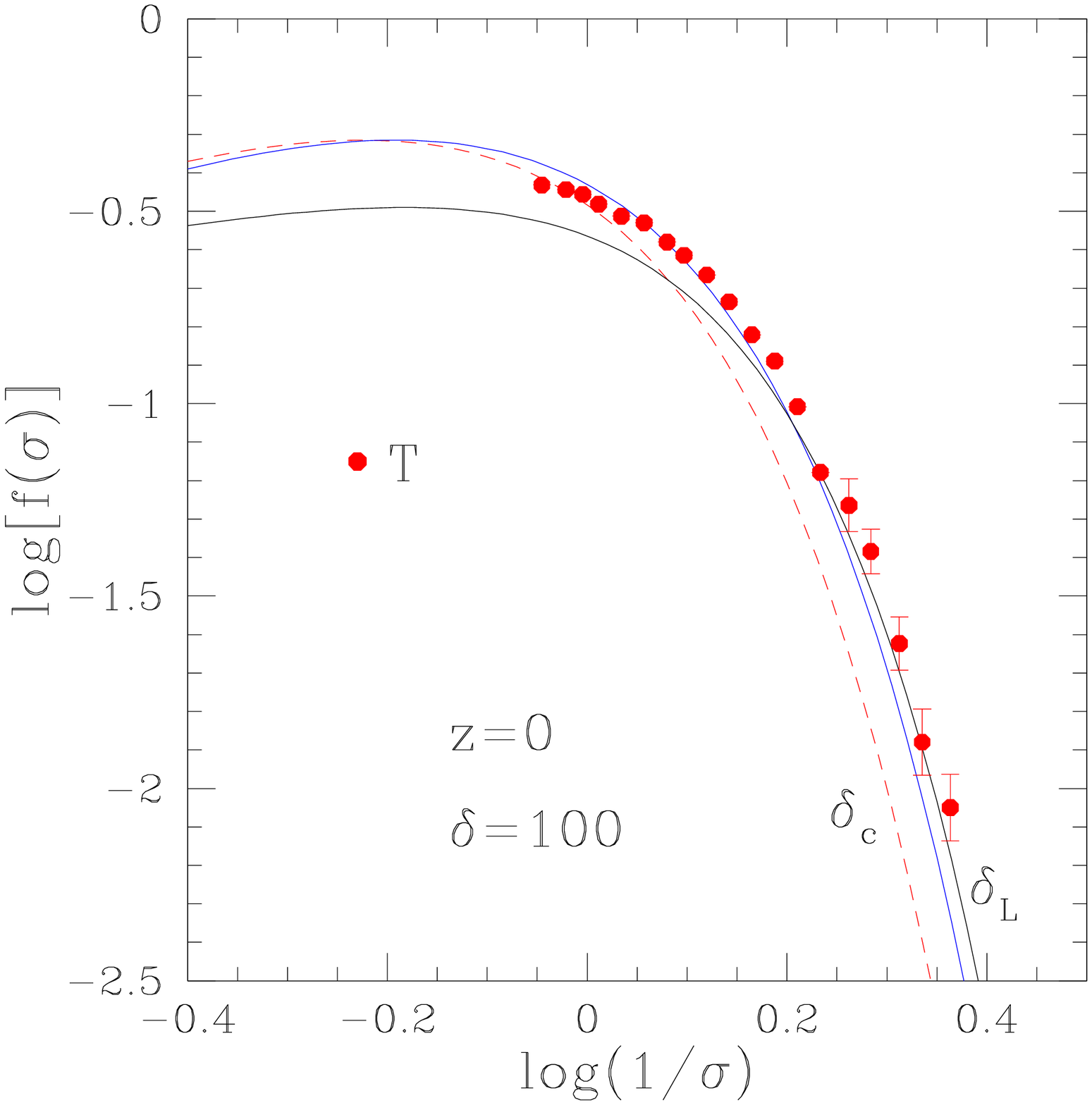}}
\epsfxsize=9 cm \epsfysize=6 cm {\epsfbox{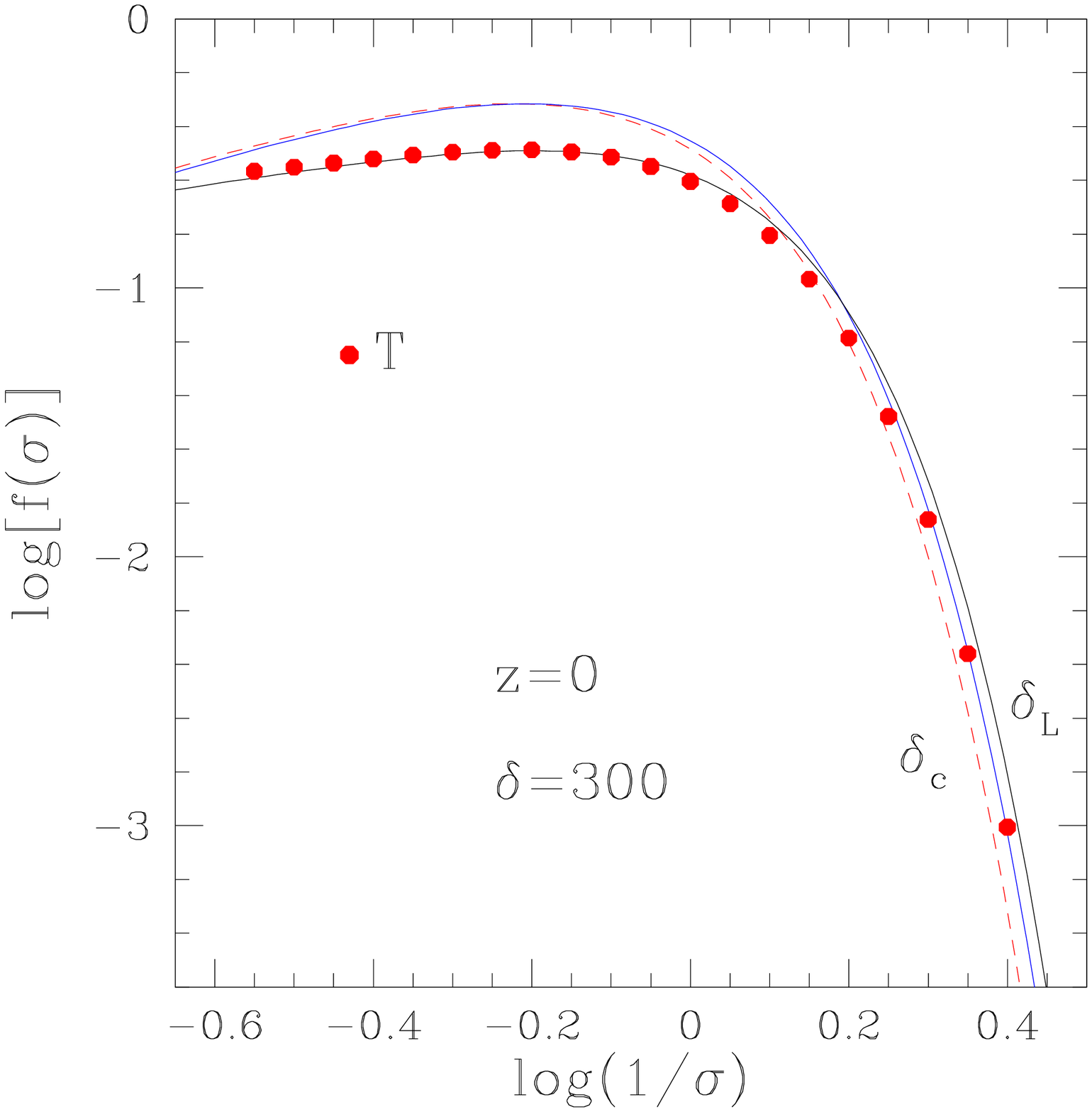}}
\end{center}
\caption{The mass functions at $z=0$ of halos defined by the nonlinear density
contrasts $\delta=100$ (upper panel) and $\delta=300$ (lower panel). 
The points show the numerical simulations of Tinker et al. (2008). 
The dashed line is the usual Press-Schechter mass function, while the solid
lines correspond to Eqs.(\ref{fPSdL}) and (\ref{fitfsig}), with now 
$\delta_L=\cF^{-1}(100)\simeq 1.55$ (upper panel) and 
$\delta_L=\cF^{-1}(300)\simeq 1.61$ (lower panel).}
\label{figfsig_z0_D100_D300}
\end{figure}

We compare in Fig.~\ref{figfsig_z0_D200} the prediction (\ref{fPSdL}) (solid
line labeled as $\delta_L=\cF^{-1}(\delta)$) with results from numerical
simulations, for the nonlinear threshold $\delta=200$ at redshift $z=0$.
In our case, this corresponds to a linear density contrast
$\delta_L \simeq 1.59$, that is obtained from Fig.~\ref{figdeltaLOm}
or the fit (\ref{deltaL200fit}).
As in section~\ref{Density-distribution}, we consider a $\Lambda$CDM cosmology
with $(\Om,\OL,\sigma_8,n_s,h)=(0.27,0.73,0.79,0.95,0.7)$.
This corresponds to the cosmological parameters of the largest-box numerical
simulations of Tinker et al. (2008), which allow the best comparison with the
theoretical predictions, as they also define halos by the density
threshold $\delta=200$ with a spherical-overdensity algorithm.
The numerical results are the fits to the mass function given in
Sheth \& Tormen (1999) (``ST''), Jenkins et al. (2001) (``J''),
Reed et al. (2003) (``R''), Warren et al. (2006) (``W'') and
Tinker et al. (2008) (``T''). Note that these mass functions are defined in
slightly different fashions, using either a spherical-overdensity or
friends-of-friends algorithm, and density contrast thresholds that vary
somewhat about $\delta=200$. However, they agree rather well, as the dependence
on $\delta$ is rather weak. This can be understood from 
Fig.~\ref{figlrho_deltaL}, which shows that around $\delta=200$ the linear
density contrast $\delta_L=\cF^{-1}(\delta)$ has a very weak dependence on
$\delta$.
We also plot the usual Press-Schechter prediction (dashed line labeled
$\delta_c$), that amounts to replace $\delta_L$ by $\delta_c=1.6754$ in
Eq.(\ref{fPSdL}) (since $\Omz=0.27$ at $z=0$). We can see that using the exact value
$\delta_L=\cF^{-1}(\delta)$ significantly improves the agreement with numerical
simulations at large masses. Note that there are no free parameters in
Eq.(\ref{fPSdL}). Of course, at small masses the mass function (\ref{fPSdL})
closely follows the usual Press-Schechter prediction and shows the same level
of disagreement with numerical simulations. This is expected since only the
exponential cutoff (\ref{nMtail}) has been exactly derived from
section~\ref{Density-distribution}. Since at large masses the power-law
prefactor $1/\sigma$ in Eq.(\ref{fPSdL}) is also unlikely to be correct,
as discussed above, we give a simple fit that matches the numerical simulations
from small to large masses (solid line that runs through the simulation points) 
while keeping the exact exponential cutoff:
\beq
f(\sigma) = 0.5 \left[ (0.6 \, \nu)^{2.5}+(0.62 \, \nu)^{0.5} \right] 
\, e^{-\nu^2/2} ,
\label{fitfsig}
\eeq
with
\beq
\nu=\frac{\delta_L}{\sigma} , \;\;\; \delta_L=\cF^{-1}(\delta) \simeq 1.59 
\;\; \mbox{for} \;\; \delta=200 , \; \Omz=0.27
\label{nudef}
\eeq
At higher redshift or for other $\Omz$ one simply needs to use the relevant
mapping $\cF^{-1}(\delta)$, shown in Fig.~\ref{figdeltaLOm} or given by the
simple fit (\ref{deltaL200fit}) for $\delta=200$.
This mass function also satisfies the normalization (\ref{normf}),
whatever the value of the threshold $\delta_L$.

Thus, we suggest that mass functions of virialized halos should be
defined with a fixed nonlinear density threshold, such as $\delta=200$,
and fits to numerical simulations should use the exact exponential cutoff
of Eq.(\ref{nMtail}), with the appropriate linear density contrast 
$\delta_L=\cF^{-1}(\delta)$, rather than treating this as a free parameter.
This would automatically ensure that the large mass tail has the right form
(up to subleading terms such as power-law prefactors), as emphasized by the
reasonable agreement with simulations of Eq.(\ref{fPSdL}) at large masses. 
Moreover, it is best no to introduce unnecessary free parameters that
become partly degenerate. Here we must note that Barkana (2004) had already
noticed that, taking the spherical collapse at face value and defining halos
by a nonlinear threshold $\delta$ (he chose $\delta=18\pi^2$), one should
use the linear threshold $\delta_L=\cF^{-1}(\delta)$ for the Press-Schechter
mass function. Unfortunately, noticing that the value of $\delta_L$ given
by fits to numerical simulations was even lower, he concluded that this was
not sufficient to reconcile theoretical predictions with numerical results.
As shown by Fig.~\ref{figfsig_z0_D200}, this is not the case, as the
parameter $\delta_L$ used in fitting formulae is partly degenerate with
the exponents of the power-law prefactors, so that it is possible to match
numerical simulations while satisfying the large-mass tail (\ref{nMtail}).
Of course, the actual justification of the asymptote (\ref{nMtail}) is
provided by the analysis of section~\ref{Density-distribution}, which
shows that below the upper bound $\delta_+$ the spherical collapse is indeed
relevant and asymptotically correct at large masses, as it corresponds to
the saddle-point of the action (\ref{Sdef}).

Note that the result (\ref{nMtail}) also implies that the mass function
$f(\sigma)$ is not exactly universal, since the mapping 
$\delta\mapsto\delta_L=\cF^{-1}(\delta)$ shows a (very) weak dependence on
cosmological parameters, see Fig.~\ref{figdeltaLOm}.
Using the exact tail (\ref{nMtail}) should also improve the robustness of
the mass function with respect to changes of cosmological parameters and
redshifts.

Next, we compare in Fig.~\ref{figfsig_z0_D100_D300} the mass functions obtained
at $z=0$ for the density thresholds $\delta=100$ and $\delta=300$ with the
numerical simulations from Tinker et al. (2008), who also considered these
density thresholds. We show the usual Press-Schechter mass function (dashed line)
and the results obtained from Eqs.(\ref{fPSdL}) and (\ref{fitfsig}) 
(solid lines), with now $\delta_L=\cF^{-1}(100)\simeq 1.55$ and 
$\delta_L=\cF^{-1}(300)\simeq 1.61$. We can see that the large mass tail remains
consistent with the simulations, but for the case $\delta=100$ it seems that the
shape of the mass function at low and intermediate masses is modified and 
cannot be absorbed through the rescaling of $\delta_L$. It appears to follow
Eq.(\ref{fPSdL}) rather than the fit (\ref{fitfsig}), but this is likely to
be a mere coincidence. We should note that Fig.~\ref{figshellcrossz} shows
that $\delta_+<300$ for massive halos, so that shell-crossing should be
taken into account and the tail (\ref{nMtail}) is no longer exact for
$\delta=300$, although it should still provide a reasonable approximation,
as checked in Fig.~\ref{figfsig_z0_D100_D300}. Thus, even though 
Tinker et al. (2008) also studied higher density thresholds, we do not consider
such cases here as the tail (\ref{nMtail}) no longer applies.

\section{Halo density profile}
\label{Halo-profile}

\begin{figure}[htb]
\begin{center}
\epsfxsize=4.42 cm \epsfysize=4.6 cm {\epsfbox{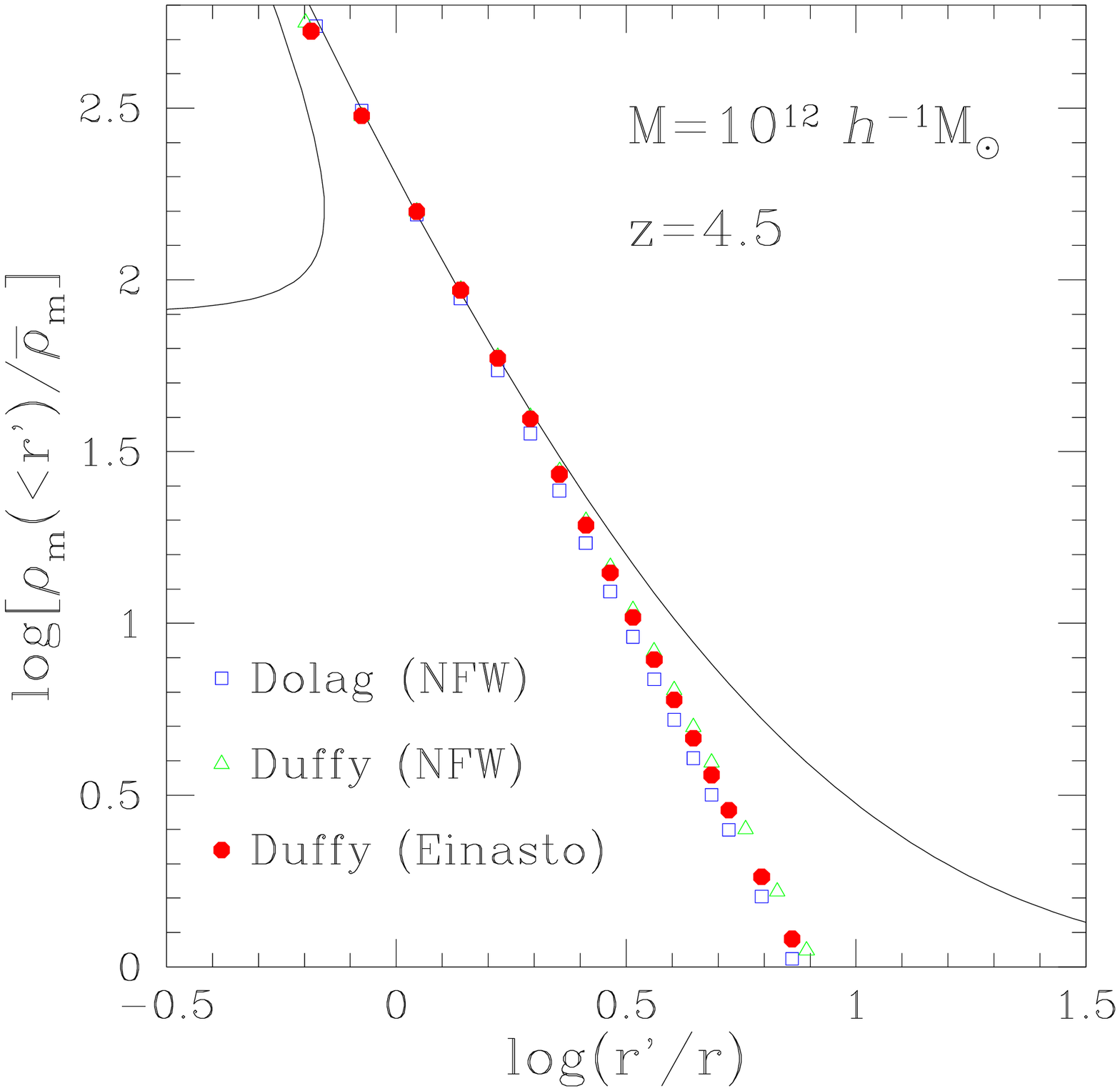}}
\epsfxsize=4.42 cm \epsfysize=4.6 cm {\epsfbox{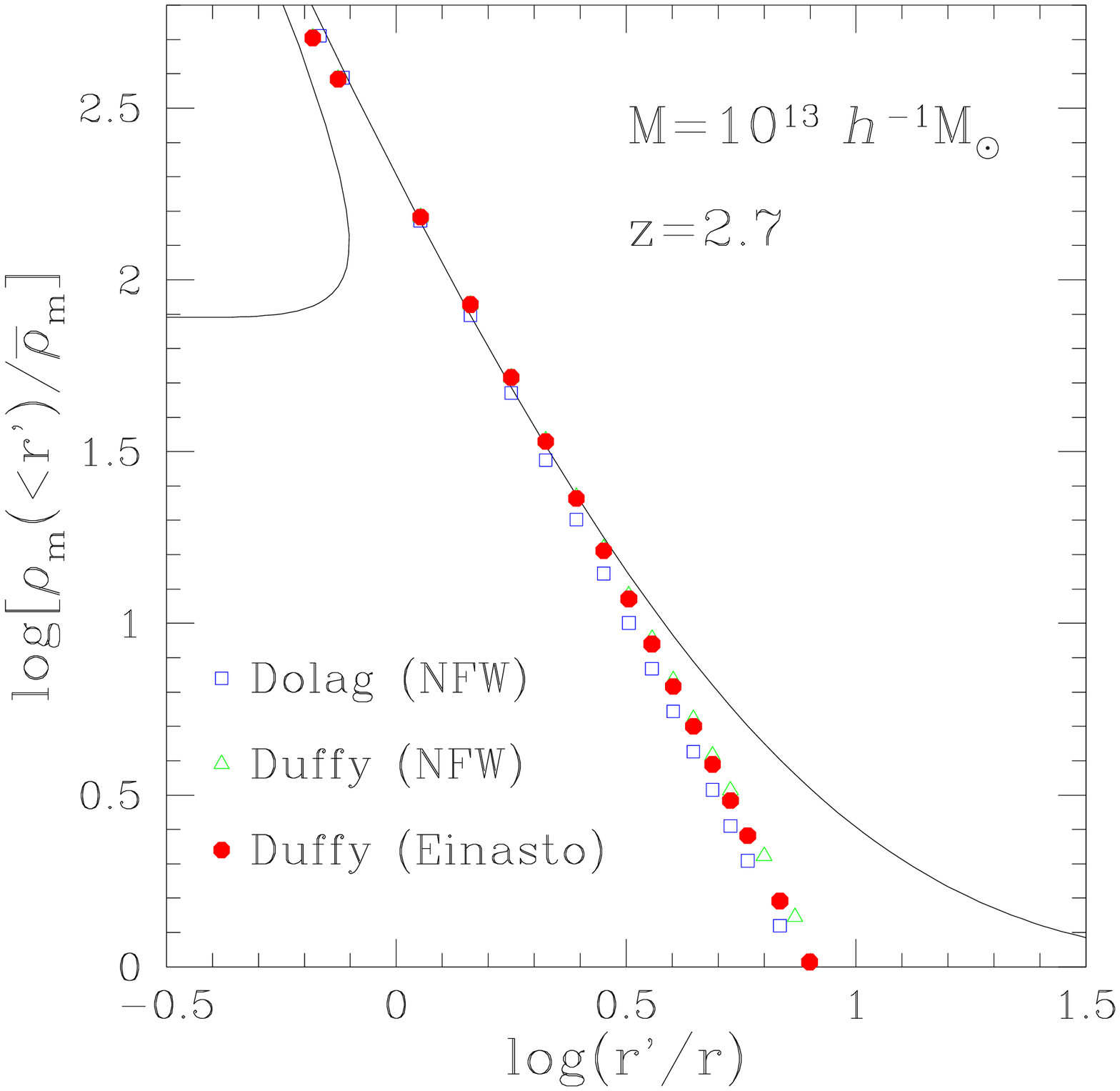}}\\
\epsfxsize=4.42 cm \epsfysize=4.6 cm {\epsfbox{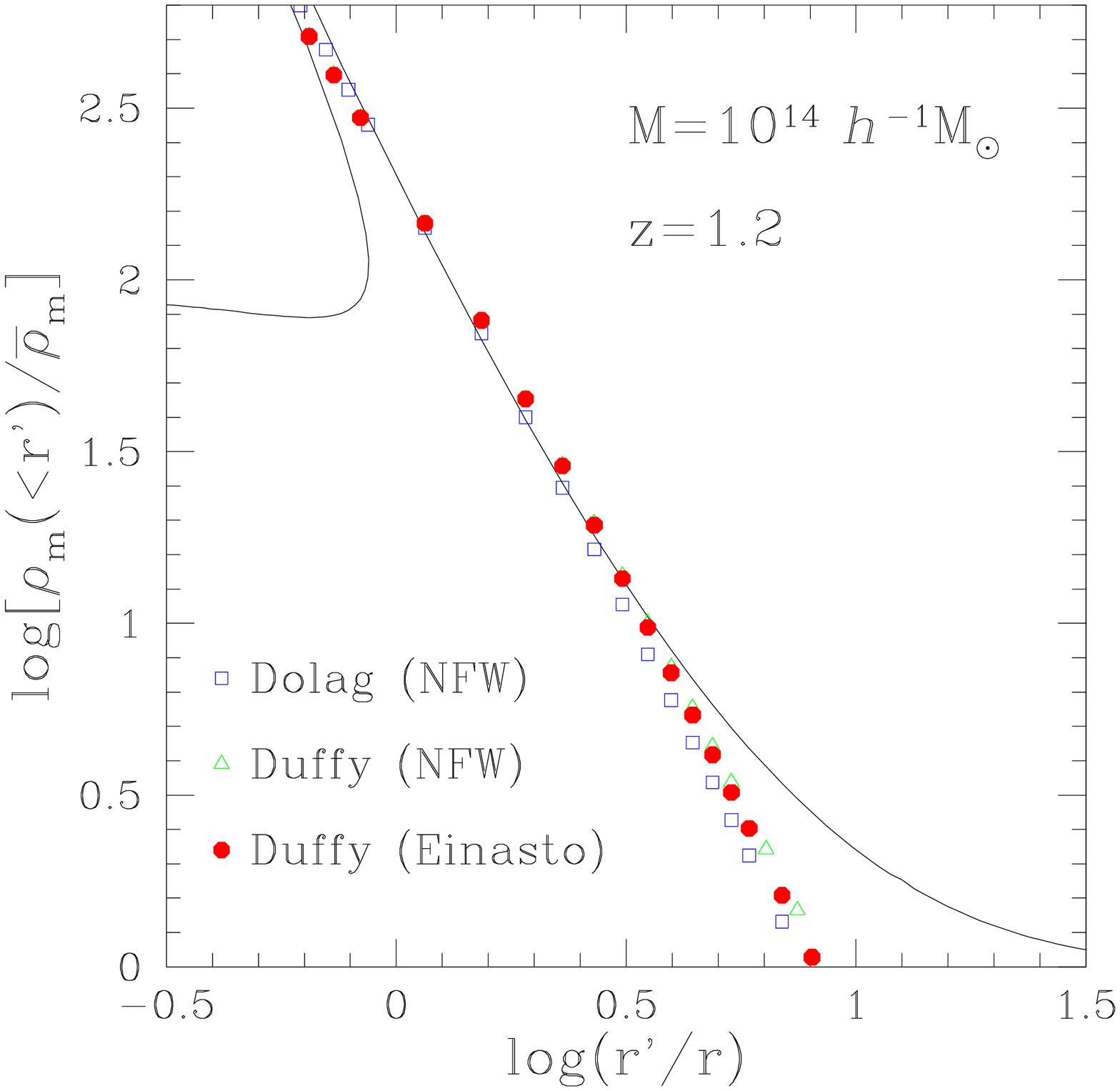}}
\epsfxsize=4.42 cm \epsfysize=4.6 cm {\epsfbox{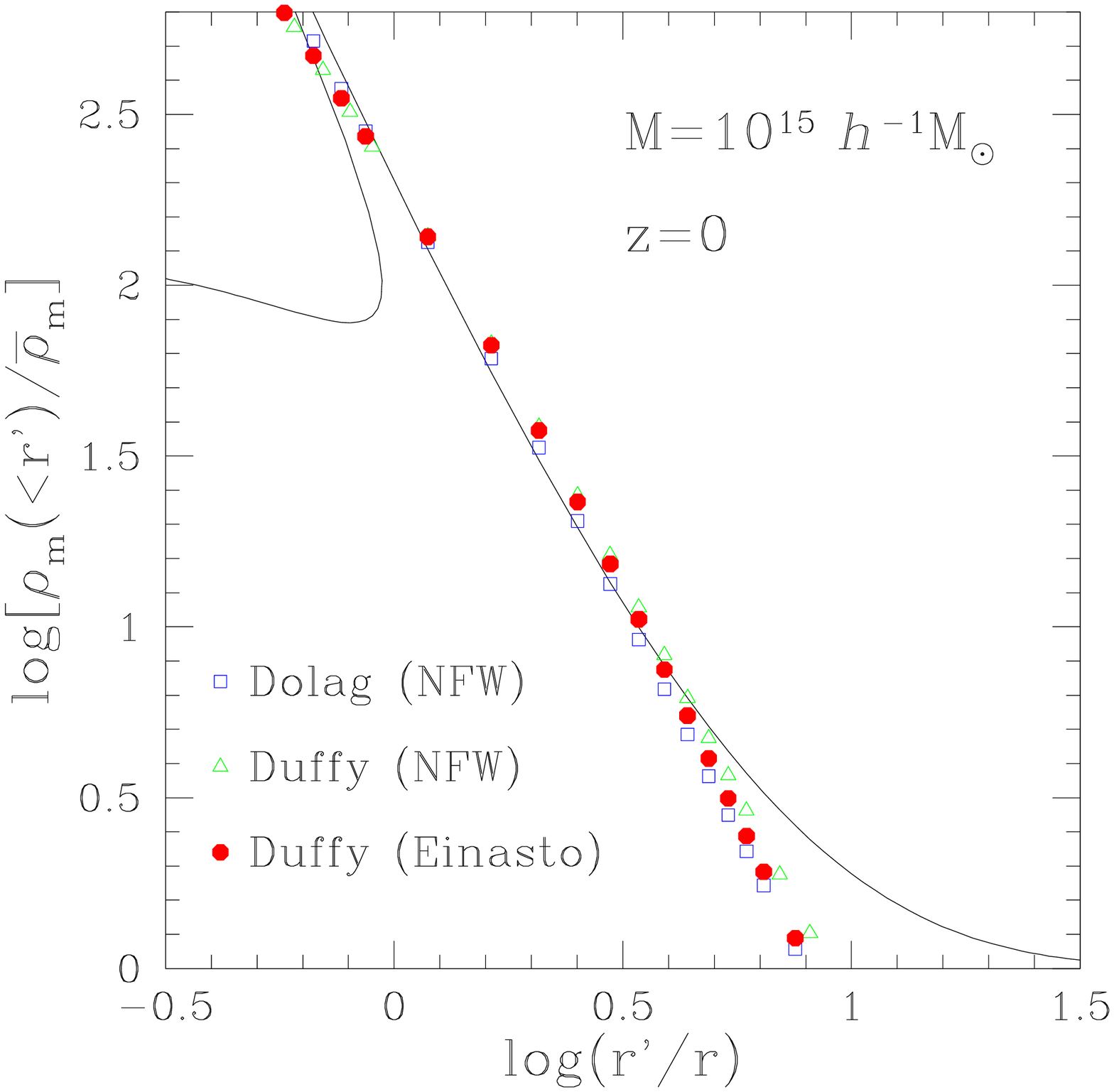}}
\end{center}
\caption{The density profile of rare massive halos, for several masses $M$.
For each mass the redshift is such that $\sigma(M,z)=0.5$, as the theoretical
prediction from Eq.(\ref{profile}) (solid line) only applies to rare events.
The second branch that appears at small radii is due to the shell-crossing,
hence the theoretical prediction only holds to the right of this branch.
The points are fits to numerical simulations, based on an NFW profile 
(Dolag et al. 2004; Duffy et al. 2008) or an Einasto profile (Duffy et al. 2008).
Note that we show the mean density within radius $r'$, $\rhom(<r')/\rhobm$,
rather than the local density at radius $r'$, so that the NFW and Einasto
profiles are integrated once.}
\label{figlrhorM}
\end{figure}

As for the mass function $n(M)$, the analysis of 
section~\ref{Density-distribution} shows that the density profile of rare
massive halos is given by the spherical saddle-point (\ref{profile}),
see also Barkana (2004) and Prada et al. (2006).
This holds for halos selected by some nonlinear density threshold $\delta$
in the limit of rare events, provided shell-crossing has not occurred beyond
the associated radius $r$. In particular, this only applies to the outer part
of the halo since in the inner part, at $r'\ll r$, shell-crossing must be
taken into account. Then, as discussed in section~\ref{Density-distribution},
a strong radial-orbit instability comes into play and modifies the profile
in this inner region, as deviations from spherical symmetry govern the dynamics
and the virialization process (Valageas 2002b).

We compare in Fig.~\ref{figlrhorM} the nonlinear density profile obtained
from Eq.(\ref{profile}) with fits to numerical simulations. We plot the
overdensity within radius $r'$, $1+\delta_r=\rhom(<r')/\rhobm$, as a function
of radius $r$. This is again obtained from Eq.(\ref{profile}) with the mapping 
$q'^3=(1+\delta_{r'})r'^3$ and $\delta_{r'}=\cF(\delta_{Lq'})$.
Since this only applies to the limit of rare events, we choose for each mass
$M$ a redshift $z$ such that $\sigma(M,z)=0.5$. Thus, smaller masses are
associated with higher redshifts. The results do not significantly
depend on the precise value of the criterium used to define rare events, 
here $\sigma(M,z)=0.5$. The points in Fig.~\ref{figlrhorM} are the results
obtained from a Navarro, Frenk \& White profile (Navarro et al. 1997, NFW),
\beq
\rho(r') = \frac{\rho_s}{(r'/r_s)(1+r'/r_s)^2} ,
\label{NFW}
\eeq
or an Einasto profile (Einasto 1965),
\beq
\ln[\rho(r')/\rho_{-2}] = - \frac{2}{\alpha} 
\left[ (r'/r_{-2})^{\alpha}-1 \right] .
\label{Einasto}
\eeq
For the NFW profile, the characteristic radius $r_s$ is obtained from the
concentration parameter, $c(M,z)=r/r_s$, where as in the previous
sections $r$ is the Eulerian radius where $\delta_r=200$ (i.e. $r=r_{200}$). 
Then, we use in Fig.~\ref{figlrhorM} the two fits obtained in 
Dolag et al. (2004) and Duffy et al. (2008) from numerical simulations,
of the form $c(M,z)= a (M/M_0)^b (1+z)^c$. For the Einasto profile,
we use the fits obtained in Duffy et al. (2008) for the concentration parameter,
$c(M,z)=r/r_{-2}$, and for the exponent $\alpha$ (written as a quadratic
polynomial over $\nu=\delta_c/\sigma(M,z)$ as in Gao et al. (2008)).
Then, we integrate over $r'$ the profiles (\ref{NFW})-(\ref{Einasto}),
to obtain the overdensity profiles $\rhom(<r')/\rhobm$ shown in
Fig.~\ref{figlrhorM} (as $\rhom(<r')$ is the mean density within radius $r'$
and not the local density at radius $r'$ as in Eqs.(\ref{NFW})-(\ref{Einasto})).

We can check in Fig.~\ref{figlrhorM} that our results agree reasonably well
with these fits to numerical simulations over the range where both predictions
are valid. Note that our prediction has no free parameter, since it is given
by the saddle-point profile (\ref{profile}).
At large radii we recover the mean
density of the Universe, while the numerical profiles 
(\ref{NFW})-(\ref{Einasto}) go to zero, but this is an artefact of the
forms (\ref{NFW})-(\ref{Einasto}), that were designed to give a sharp boundary
for the halos and are mostly used for the high-density regions.
For a study of numerical simulations at outer radii, see Figs.8 and 10 of
Prada et al. (2006) which recover the mean density of the Universe at large
scales. 
At small radii, the second branch that makes a turn somewhat below $r$ is
due to shell-crossing that makes the function $\rhom(<r')$ bivaluate. Below
the maximum shell-crossing radius, for each Eulerian radius $r'$ there are
two Lagrangian radii $q'$, a large one that corresponds to shells that are
still falling in, and a smaller one that corresponds to shells that have
already gone once through the center (note that in agreement with 
Fig.~\ref{figrq}, shell-crossing appears at slightly smaller relative radii
for smaller mass). 
Then, the theoretical prediction only holds to the right of this second branch,
where there is only one branch and no shell-crossing.
Note that the theoretical prediction (\ref{profile}) explicitly shows that the
halo density profiles are not universal. Within the phenomenological fits
(\ref{NFW})-(\ref{Einasto}) this is parameterized through the dependence
on mass and redshift of the concentration parameter and of the exponent
$\alpha$. However, we can see that over the regime where Eq.(\ref{profile})
applies the local slope of the halo density is $\rho(r') \sim r'^{-2}$, which
explains the validity of the fits (\ref{NFW})-(\ref{Einasto}) in this domain.
Unfortunately, our approach cannot shed light on the inner density profile,
where $\delta_{r'}\gg 200$, which is the region of interest for most
practical applications of the fits (\ref{NFW})-(\ref{Einasto}).

The same approach, based on the spherical collapse, was studied in greater
details in Betancort-Rijo et al. (2006) and Prada et al. (2006).
They consider the ``typical'' profile (\ref{profile}) that we study here,
as well as ``mean'' and ``most probable'' profiles. In agreement with the
steepest-descent approach of section~\ref{Density-distribution}, they
find that the most probable profile closely follows the typical profile
(\ref{profile}) and they obtain a good match with numerical simulations
for massive halos, paying particular attention to outer radii.
Therefore, we do not further discuss halo profiles here.

\section{Halo bias}
\label{Halo-bias}

In addition to their multiplicity and their density profile, a key property
of virialized halos is their two-point correlation function. At large scales
it is usually proportional to the matter density correlation, up to a
multiplicative factor $b^2$, called the bias of the specific halo population.
We revisit in this section the derivation of the bias of massive halos,
following Kaiser (1984), and we point out that 
paying attention to some details it is possible to reconcile the theoretical
predictions with numerical simulations, without introducing any free parameter.

As seen in the previous sections, rare massive halos can be identified with
rare spherical fluctuations in the initial (linear) density field.
More precisely, as in section~\ref{Density-distribution} we may consider 
the bivariate density distribution, $\cP(\delta_{r_1},\delta_{r_2})$, of the 
density contrasts $\delta_{r_1},\delta_{r_2}$, in the spheres of radii $r_1,r_2$,
centered at points $\bx_1,\bx_2$. Thus, we introduce as in Eq.(\ref{phidef})
the double Laplace transform $\varphi(y_1,y_2)$,
\beqa
e^{-\varphi(y_1,y_2)/\sigma^2} & = &
\lag e^{-(y_1\delta_{r_1}+y_2\delta_{r_2})/\sigma^2} \rag \nonumber \\
&& \hspace{-1.3cm} = \int_{-1}^{\infty} \dd\delta_{r_1} \dd\delta_{r_2} \, 
e^{-(y_1\delta_{r_1}+y_2\delta_{r_2})/\sigma^2} \,
\cP(\delta_{r_1},\delta_{r_2}) ,
\label{phi12def}
\eeqa
where $\sigma^2=\sigma^2_{r,r}(0)$ is the linear variance (\ref{sigr1r2})
at some scale $r$. If $r_1=r_2$ we may take $r=r_1=r_2$, otherwise it can be
any intermediate scale, as the results do not depend on this factor.
The only requirement is that $\sigma$ should scale as $\sigma_{r_1}$ and
$\sigma_{r_2}$, which is obtained by choosing for instance a fixed ratio
$r/\sqrt{r_1 r_2}$. 
Then, the probability distribution $\cP(\delta_{r_1},\delta_{r_2})$ can be
obtained from the cumulant generating function $\varphi(y_1,y_2)$ through
a double inverse Laplace transform, as in Eq.(\ref{Pphi}). Next,
$\varphi(y_1,y_2)$ can be written in terms of the linear density field
as a path integral, such as (\ref{path}), with an action that now reads as
\beq
\cS[\delta_L] = y_1 \, \delta_{r_1}[\delta_L] + y_2 \, \delta_{r_2}[\delta_L] 
+ \frac{\sigma^2}{2} \delta_L . C_L^{-1} . \delta_L
\label{S12def}
\eeq
Then, for rare events the tail of the distribution 
$\cP(\delta_{r_1},\delta_{r_2})$ is governed by the last Gaussian weight
of (\ref{S12def}), as in Eq.(\ref{PdL}),
\beq
\sigma \rightarrow 0 : \;\; \cP(\delta_{r_1},\delta_{r_2}) \sim 
e^{-\frac{1}{2} \delta_L . C_L^{-1} . \delta_L} ,
\label{Ptail12}
\eeq
where $\delta_L[\bq]$ is the relevant saddle-point of the action $\cS$.
Unfortunately, since the action (\ref{S12def}) is no longer spherically
symmetric, it is not possible to obtain an explicit expression of its
minimum. Therefore, we must rely on some simple approximations.
In the limit of large distance, $x_{12}=|\bx_2-\bx_1|\rightarrow\infty$,
between the positions of both halos, the linear field $\delta_L(\bs)$
around the Lagrangian positions $\bs_1,\bs_2$, of the halos should follow
the profile (\ref{profile}) (we note the Lagrangian coordinates $\bs_i$ to
avoid confusion with the Lagrangian radii $q_i$).
Thus, we neglect the tidal forces of the halos, but we keep
track of the mean displacement of each halo, due to the gravitational attraction
from the other one, by distinguishing their Lagrangian positions $\bs_1,\bs_2$,
from their Eulerian positions $\bx_1,\bx_2$.
Within this approximation, the distribution $\cP(\delta_{r_1},\delta_{r_2})$
can be estimated from the Gaussian distribution 
$\cP_L(\delta_{Lq_1},\delta_{Lq_2})$
of the linear density contrasts $\delta_{Lq_1},\delta_{Lq_2}$, at positions
$\bs_1,\bs_2$, within the Lagrangian spheres of radii $q_1$ and $q_2$,
with the mapping (\ref{qr}). This closely follows the approach introduced in
Kaiser (1984), except for the distinction between $\bs_i$ and $\bx_i$.
The joint distribution $\cP_L(\delta_{Lq_1},\delta_{Lq_2})$ reads as
(see also Kaiser 1984; Politzer \& Wise 1984)
\beq
\cP_L(\delta_{Lq_1},\delta_{Lq_2}) =  \frac{1}{(2\pi)\sqrt{\det M}} \, 
e^{-\frac{1}{2} \sum_{i,j} \delta_{Lq_i} . M^{-1}_{ij} . \delta_{Lq_j}} ,
\label{PL12}
\eeq
where $M$ is the linear covariance matrix,
\beq
M = \left( \bea{cc} \sigma^2_1 \, & \, \sigma^2_{12} \\ & \\
\sigma^2_{12} \, & \, \sigma^2_2 \ea \right) .
\label{Mdef}
\eeq
Here we defined from Eq.(\ref{sigr1r2}), $\sigma^2_i=\sigma^2_{q_i,q_i}(0)$
and $\sigma^2_{12}=\sigma^2_{q_1,q_2}(s_{12})$ with $s_{12}=|\bs_2-\bs_1|$.
The inverse of the matrix $M$ writes as
\beq
M^{-1} =  \frac{1}{\sigma^2_1\,\sigma^2_2-\sigma^4_{12}} 
\left( \bea{cc} \sigma^2_2 & -\sigma^2_{12} \\ & \\
-\sigma^2_{12} & \sigma^2_1 \ea \right) .
\label{Minv}
\eeq
This yields
\beqa
\cP_L(\delta_{Lq_1},\delta_{Lq_2}) & = & \cP_L(\delta_{Lq_1}) \, 
\cP_L(\delta_{Lq_2}) \, \frac{\sigma_1\sigma_2}
{\sqrt{\sigma^2_1\,\sigma^2_2-\sigma^4_{12}}} \nonumber \\
&& \hspace{-2.3cm} \times
\, \exp \!\left[ \frac{2\delta_{Lq_1}\delta_{Lq_2}\sigma^2_{12}
-\delta_{Lq_1}^2\sigma^4_{12}/\sigma^2_1
-\delta_{Lq_2}^2\sigma^4_{12}/\sigma^2_2}
{2(\sigma^2_1\,\sigma^2_2-\sigma^4_{12})} \right] .
\label{PL12_PL1PL2}
\eeqa
Next, defining the real-space halo correlation $\xi_{M_1,M_2}(r)$ as the
fractional excess of halo pairs (Kaiser 1984; Peebles 1980), 
\beqa
n_{\bx_1,\bx_2}(M_1,M_2)\dd M_1\dd M_2 \dd\bx_1\dd\bx_2 & = & \nonumber \\
&& \hspace{-4.7cm} \nb(M_1) \nb(M_2) \left(1+\xi_{M_1,M_2}(x_{12})\right)
\dd M_1\dd M_2 \dd\bx_1\dd\bx_2 ,
\eeqa
which also gives the conditional probability,
\beq
n_{\bx_1,\bx_2}(M_2|M_1) = \nb(M_2) \left(1+\xi_{M_1,M_2}(x_{12})\right) ,
\eeq
we write
\beqa
1+\xi_{M_1,M_2}(x_{12}) & \sim & (1+\delta_M(x_{12})) 
\frac{\cP_L(\delta_{L1},\delta_{L2})}{\cP_L(\delta_{L1}) \cP_L(\delta_{L2})} ,
\label{xiM1M2def}
\eeqa
where the mass of each halo is given by $M_i=\rhobm4\pi q_i^3/3$ and
$\delta_{Li}=\cF^{-1}(\delta_i)$ are the linear density contrasts which are
associated with the nonlinear density contrasts $\delta_i$ that define the
halos, as in section~\ref{Mass-function} and Fig.~\ref{figdeltaLOm}.
In Eq.(\ref{xiM1M2def}) the factor $(1+\delta_M(x_{12}))$ models the effects
associated with the mapping from Lagrangian to Eulerian space.
Indeed, in the limit of rare massive halos and large separation, the number
of neighbors is conserved and we write $n_{\rm Eul}x_{12}^2\dd x_{12}
= n_{\rm Lag}s_{12}^2\dd s_{12}$ (for an integral form of the
conservation of pairs, or neighbors, see Peebles 1980).
We take $\dd\bs_{12}=(1+\delta_M)\dd\bx_{12}$, where we define
$\delta_M(x_{12})$ as the local nonlinear density contrast at Eulerian
distance $x_{12}$ from a halo of mass $M$, obtained from the profile 
(\ref{profile}), and we choose $M=\max(M_1,M_2)$ (which obeys the symmetry 
$M_1\leftrightarrow M_2$), whence $q=\max(q_1,q_2)$.
This reflects the fact that the gravitational attraction of a massive halo pulls
matter towards it center with a strength that depends on distance, so
that the Jacobian $|\pl\bq/\pl\bx|=(1+\delta)$ is different from
unity. In a sense this is a tidal effect, as a locally volume-preserving
displacement would not affect the number densities, which is why we take for
$\delta_M(x_{12})$ the density contrast at distance $x_{12}$ and not the density
contrast within radius $x_{12}$.
We may note that a local bias model (Mo \& White 1996), using a peak-background
split argument (Efstathiou et al. 1988), would rather give a quadratic factor
$(1+\delta_1)(1+\delta_2)$. However, we prefer to keep the linear factor
(\ref{xiM1M2def}) as it also remains consistent in case of large 
displacements\footnote{For instance, let us consider a large system which can
be subdivided into cells of two classes, $\{+,-\}$, with matter 
densities $\{\rho_+,\rho_-\}$ and volume fractions  $\{\eta_+,\eta_-\}$, and
$\rho_+>\rhob>\rho_-$, $\eta_+<\eta_-$. Then, from the conservation of
volume ($\eta_+ + \eta_-=1$) and mass ($\eta_+\rho_+ + \eta_-\rho_-=\rhob$),
we obtain in the limits $\rho_+\gg 1$ and $\rho_-\ll 1$, 
$\xi=\lag\rho^2\rag_c/\rhob^2 \sim \rho_+/\rhob$.}. This also expresses the fact
that all pairs cannot simultaneously get closer (as fluctuations grow
some objects move closer but they also further separate from other emerging
groups).
Then, defining the halo bias as
\beq
b^2_{M_1,M_2}(r) = \frac{\xi_{M_1,M_2}(r)}{\xi(r)} ,
\label{biasdef}
\eeq
where $r=x_{12}$ and $\xi(r)$ is the nonlinear matter correlation, 
we obtain the bias from Eq.(\ref{xiM1M2def}).
Note that the bias (\ref{biasdef}) does not factorize, 
$b^2_{M_1,M_2}(r) \neq b_{M_1}(r) b_{M_2}(r)$, because of the terms $\sigma_{12}$
and $\delta_M$.

At large separation, $r \rightarrow \infty$,
we are in the linear regime, so that the matter correlation reads as
$\xi(r) \simeq \sigma^2_{0,0}(r)$, and the local density contrast
$\delta_M(r)$ is small and close to the linear density contrast
$\delta_L(s)$. Note that this is the density contrast at Lagrangian radius
$s$, and not the mean density contrast within radius $s$. Therefore,
it is related to Eq.(\ref{profile}) by
\beq
3 q^2 \delta_L(q) = \frac{\pl}{\pl q}\left( q^3 \delta_{Lq} \right) ,
\eeq
whence
\beq
\delta_L(s) = \delta_L \, \frac{{\tilde\sigma}^2(q,s)}
{\sigma^2_q} ,
\label{profile_s12}
\eeq
with $\delta_L=\cF^{-1}(200)$ for instance, and using Eq.(\ref{sigr1r2})
\beq
{\tilde\sigma}^2(q,s) =  4\pi\int\dd k \, k^2 P_L(k) W(kq) 
\frac{\sin(ks)}{ks} = \sigma^2_{q,0}(s) .
\label{tsigmadef}
\eeq

Then, from Eq.(\ref{biasdef}), the bias of halos defined by the same linear
threshold, $\delta_L=\cF^{-1}(\delta)$, reads as 
\beqa
b^2_{M_1,M_2}(r) & \!=\! & \frac{1}{\sigma^2_{0,0}(r)} \Biggl[ -1 +
\frac{\sigma_{q_1}\sigma_{q_2} \, (1+\delta_L\,\sigma^2_{q,0}(s)/\sigma^2_{q})}
{\sqrt{\sigma^2_{q_1}\sigma^2_{q_2}-\sigma^4_{q_1,q_2}(s)}} \nonumber \\
&& \hspace{-2cm} \times \exp\!\left(\frac{2\delta_L^2\sigma^2_{q_1,q_2}(s)
\!-\!\delta_L^2\sigma^4_{q_1,q_2}(s)/\sigma^2_{q_1}\!
-\!\delta_L^2\sigma^4_{q_1,q_2}(s)/\sigma^2_{q_2}}
{2[\sigma^2_{q_1}\sigma^2_{q_2}-\sigma^4_{q_1,q_2}(s)]} \right) \Biggl] .
\nonumber \\
&& 
\label{b2M1M2}
\eeqa
For equal-mass halos this simplifies as
\beqa
b^2_M(r) & = & \frac{1}{\sigma^2_{0,0}(r)} \Biggl[ -1 +
\left(\!1\!+\!\delta_L\frac{\sigma^2_{q,0}(s)}{\sigma^2_{q}}\!\right) 
\frac{\sigma^2_q}{\sqrt{\sigma^4_q-\sigma^4_{q,q}(s)}} \nonumber \\
&& \hspace{0.8cm} \times \exp\left(\frac{\delta_L^2\sigma^2_{q,q}(s)
-\delta_L^2\sigma^4_{q,q}(s)/\sigma^2_q}
{\sigma^4_q-\sigma^4_{q,q}(s)} \right) \Biggl] .
\label{b2M}
\eeqa
Note that the argument of the exponential is not necessarily small,
as stressed in Politzer \& Wise (1984). Indeed, we only assumed a large
separation limit, i.e. $\sigma^2_{q,q}(s)\ll \sigma^2_q$, and a rare-event
limit, $\delta_L/\sigma_q \gg 1$. Thus, at fixed (small) ratio
$\sigma^2_{q,q}(s)/\sigma^2_q$, we obtain a large exponent in the limit
of large-mass halos, $\sigma^2_q \rightarrow 0$. Then, keeping the full
expressions (\ref{b2M1M2}) or (\ref{b2M}) gives a nonlinear bias, since
$b^2_M(r)$ is not a simple number and shows a non-trivial scale dependence,
as will be clearly seen in Fig.~\ref{figbiasr_z10} below.
Indeed, these results cannot be recovered through a linear biasing scheme,
where the fluctuations of the halo number density field, 
$\delta_{\rm halo}= (n-\nb)/\nb$, are written as 
$\delta_{\rm halo}(M)=b(M) \delta_R$ (where the matter density field
is smoothed over some larger scale $R$), which would lead to
$\xi_M(r) \propto \sigma^2_{R,R}(r)$ whereas Eqs.(\ref{b2M1M2})-(\ref{b2M})
generate powers of all orders over $\sigma^2_{q,q}(r)$. 
In the local bias framework (Fry \& Gaztanaga 1993; Mo et al. 1997), 
where one writes $\delta_{\rm halo}(M)= \sum_i b_i(M) \delta^i_R$,
this could be interpreted as non-zero bias coefficients $b_i$ for $i \geq 2$.
However, Eqs.(\ref{b2M1M2})-(\ref{b2M}) are not equivalent to such a local
model, inspired from a peak-background split argument (Efstathiou et al. 1988).
Indeed, they do not involve any external smoothing scale $R$ and
they depend on the three linear correlations $\sigma^2_{0,0}(r), 
\sigma^2_{q,0}(r)$ and $\sigma^2_{q,q}(r)$.

Finally, we must express the
Lagrangian separation $s$ in terms of the Eulerian distance $r$.
At lowest order we again consider each halo as a test particle that falls into
the potential well built by the other halo (i.e. we neglect backreaction
effects). Then, from the analysis of section~\ref{Density-distribution} and the
linear profile (\ref{profile}), we know that a test particle at Lagrangian
distance $q'$ from one halo has moved to position $r'$, according to the
mapping (\ref{qr}). Using Eq.(\ref{profile}) this gives at first order
\beq
q'\simeq r' \left(\!1+\frac{\delta_{r'}}{3}\!\right) 
\simeq r' \left(\!1+\frac{\delta_{Lq'}}{3}\!\right) ,
\eeq
since at large distance we have $\delta_{r'} \simeq \delta_{Lq'} \ll 1$.
Therefore, we obtain the Lagrangian separation $s$ as
the solution of the implicit equation
\beq
r= s \left( 1 - \frac{\delta_L}{3} \, \frac{\sigma^2_{q_1,s}}{\sigma^2_{q_1}}
- \frac{\delta_L}{3} \, \frac{\sigma^2_{q_2,s}}{\sigma^2_{q_2}}   \right),
\label{sr}
\eeq
where again we only kept the first-order term and we took into account the
displacements of both halos.
Together with Eq.(\ref{b2M1M2}), or Eq.(\ref{b2M}), this defines our
prediction for the bias of massive collapsed halos.
Note that this approach also applies to the cross-correlation between different
redshifts.

At large separation, $r\rightarrow \infty$, and fixed mass
(i.e. fixed $\sigma_q$), we may linearize the bias 
(\ref{b2M}) over $\sigma^2(s)$ as
\beq
r\rightarrow \infty : \;\;  b^2_M(r) \sim \frac{1}{\sigma^2_{0,0}(r)} 
\left[ \delta_L \frac{\sigma^2_{q,0}(s)}{\sigma^2_q} 
+ \delta_L^2 \frac{\sigma^2_{q,q}(s)}{\sigma^4_q} \right] .
\label{bLin}
\eeq
For large masses, where $\sigma_q\ll 1$, but not too large, so that
the exponent in (\ref{b2M}) is still small, this gives
\beq
r\rightarrow \infty, \;\; M \rightarrow \infty : \;\;\; 
b_M(r) \sim \frac{\delta_L}{\sigma^2_q} \, 
\frac{\sigma_{q,q}(s)}{\sigma_{0,0}(r)} .
\label{bLinM}
\eeq
Thus we recover the result of Kaiser (1984) and Mo \& White (1996), except for
the multiplicative factor $\sigma_{q,q}(s)/\sigma_{0,0}(r)$. It expresses the
facts that halos only probe the linear density field smoothed over the
Lagrangian scale $q$ (i.e. the formation of a halo does not depend on
wavelengths much smaller than its radius) and that halos have moved from
distance $s$ to $r$ by their mutual gravitational attraction.
This factor yields a weak scale dependence for $b(M)$, as the slope of the
linear power spectrum slowly varies with scale $r$.
We can also note that at lowest order over $\sigma^2(s)$ we may write the
solution of Eq.(\ref{sr}) as
\beq
s= r \left( 1 + \frac{\delta_L}{3} \, \frac{\sigma^2_{q_1,r}}{\sigma^2_{q_1}} 
+ \frac{\delta_L}{3} \, \frac{\sigma^2_{q_2,r}}{\sigma^2_{q_2}} \right) ,
\label{srLin}
\eeq
which provides a simple explicit expression for $s$.

\begin{figure}[htb]
\begin{center}
\epsfxsize=9 cm \epsfysize=7 cm {\epsfbox{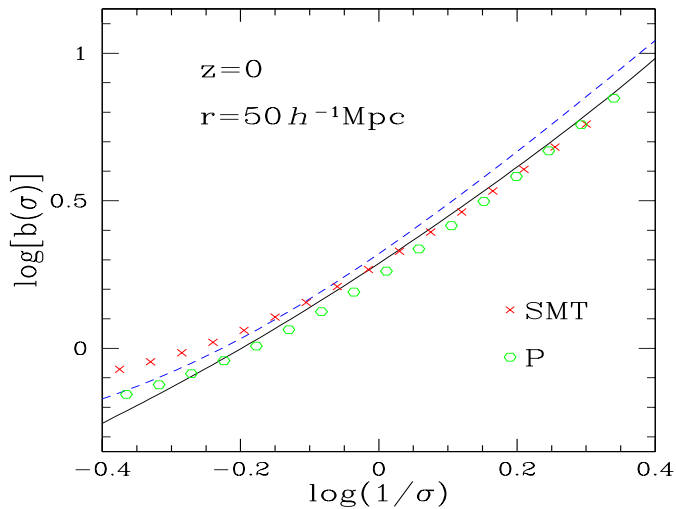}}
\end{center}
\caption{The halo bias $b(\sigma)$, as a function of $\sigma(M)$, at redshift
$z=0$ and distance $r=50 h^{-1}$Mpc. The solid line is the theoretical prediction
(\ref{b2M})-(\ref{sr}), and the dashed line is the bias obtained in
Mo \& White (1996). The points are the fits to numerical simulations,
from Sheth, Mo \& Tormen (2001) (crosses) and Pillepich (2009) (circles).}
\label{figbiasM_z0}
\end{figure}

\begin{figure}[htb]
\begin{center}
\epsfxsize=9 cm \epsfysize=7 cm {\epsfbox{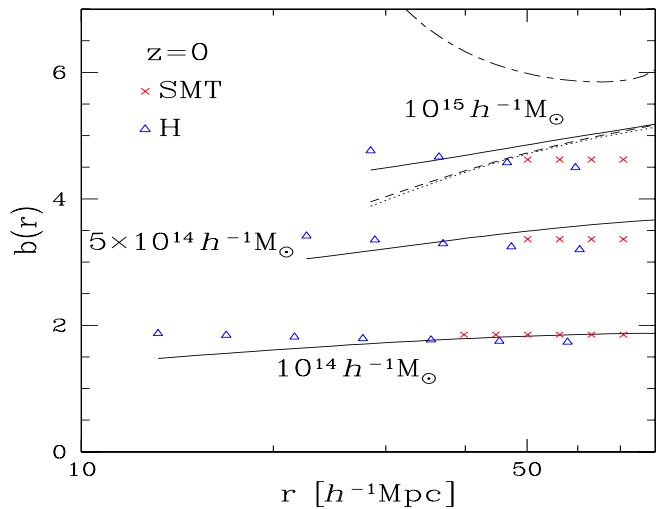}}
\end{center}
\caption{The halo bias $b(r)$, as a function of scale $r$, at redshift $z=0$ and
for several masses. The solid line is the theoretical prediction
(\ref{b2M})-(\ref{sr}). The crosses show the large-scale fit to numerical
simulations from Sheth, Mo \& Tormen (2001), while the triangles show the fit
from Hamana et al. (2001). The dashed line is the linearized bias (\ref{bLin}),
the dot-dashed line is the nonlinear bias (\ref{b2M}) where we set
$s=r$ while the dotted line uses Eq.(\ref{srLin}) 
(for $M=10^{15}h^{-1}M_{\odot}$ in the three cases).
We only plot our predictions for $r\geq 2q$.}
\label{figbiasr_z0}
\end{figure}

We compare in Figs.~\ref{figbiasM_z0}-\ref{figbiasr_z10} the bias obtained from
Eqs.~(\ref{b2M}), (\ref{sr}), with fits to numerical simulations.
We first show in Fig.~\ref{figbiasM_z0} our prediction as a function of halo
mass, at redshift $z=0$ and distance $r=50 h^{-1}$ Mpc. This is typically the
scale that is considered in numerical simulations to compute the large-scale
bias, as $b(r)$ is expected to be almost constant at large scales 
(Kaiser 1984; Mo \& White 1996), see also Eq.(\ref{bLinM}).
We also plot the standard theoretical prediction from Mo \& White (1996) 
(dashed line). We can see that our result (\ref{b2M})-(\ref{sr}) agrees rather
well with numerical simulations and the popular fit from
Sheth, Mo \& Tormen (2001).
As expected, it follows the trend of the prediction from Mo \& White (1996),
since both derivations follow the spirit of Kaiser (1984) (i.e. one identifies
halos from overdensities in the linear density field) and they agree at large
scale for rare massive halos, up to a factor of order unity, as seen in 
Eq.(\ref{bLinM}). Note that Eqs.~(\ref{b2M})-(\ref{sr}) only apply to the 
rare-event limit, as for small objects the approximations used in the derivation
no longer apply. In particular, halos can no longer be considered as spherical
isolated objects, and one should take into account merging effects.
Note that this caveat also applies to other analytical approaches, such as
Kaiser (1984) and  Mo \& White (1996). Then, our prediction should only be
used for large masses, for instance such that $b>1$.

\begin{figure}[htb]
\begin{center}
\epsfxsize=9 cm \epsfysize=7 cm {\epsfbox{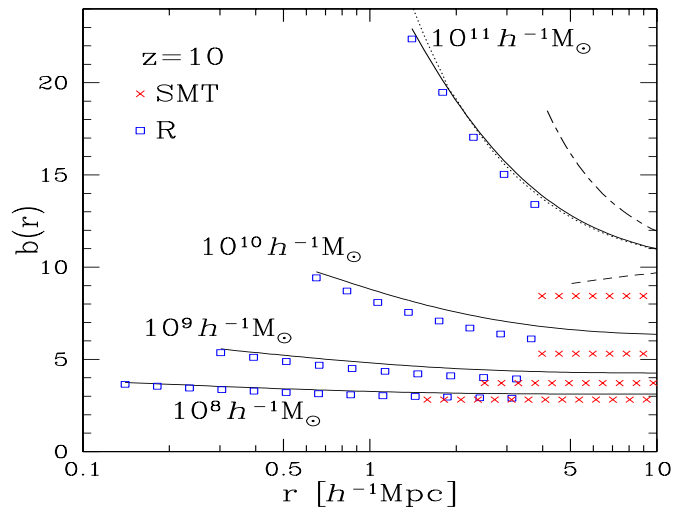}}
\end{center}
\caption{The halo bias $b(r)$, as a function of scale $r$, at redshift $z=10$
and for several masses. As in Fig.~\ref{figbiasr_z0}, the solid line is the
prediction (\ref{b2M})-(\ref{sr}), while the crosses show the large-scale fit
from Sheth, Mo \& Tormen (2001), but the squares now show the fit to numerical
simulations from Reed et al. (2009). The dashed line is the linearized bias
(\ref{bLin}), the dot-dashed line is the nonlinear bias (\ref{b2M})
where we set $s=r$ while the dotted line uses Eq.(\ref{srLin}) 
(for $M=10^{11}h^{-1}M_{\odot}$ in the three cases). Again we only plot
our predictions for $r\geq 2q$.} 
\label{figbiasr_z10}
\end{figure}

Next, we compare in Fig.~\ref{figbiasr_z0} the dependence on $r$ of the
bias (\ref{b2M})-(\ref{sr}) with the fit to numerical simulations from
Hamana et al. (2001) (using their cosmological parameters).
The crosses are the large-scale limit given by Sheth, Mo \& Tormen (2001)
and we only plot our prediction (solid lines) down to scale $r=2q$,
since it should only apply to large halo separations.
We can see that the scale-dependence that we obtain is opposite to the one
observed in the simulations. However, both are very weak and the prediction
(\ref{b2M})-(\ref{sr}) may still lie within error bars of numerical results.
For the largest mass, $M=10^{15}h^{-1}M_{\odot}$, we also plot for illustration
the linearized bias (\ref{bLin}) (lower dashed line), and the nonlinear bias
(\ref{b2M}) where we set $s=r$ (upper dot-dashed line) or we use 
Eq.(\ref{srLin}) (lower dotted line).
As expected, at very large scales the linearized bias (\ref{bLin}) agrees with
the nonlinear expression (\ref{b2M}). Using the simpler Eq.(\ref{srLin})
also gives the same results at large scales, but the Lagrangian to Eulerian
mapping still gives a non-negligible correction as shown by the upper
dot-dashed line where we set $s=r$. In any case, it is always best 
to use the full expression (\ref{b2M})-(\ref{sr}).

We compare in Fig.~\ref{figbiasr_z10} our results at high redshift,
$z=10$, with the fit to numerical simulations from Reed et al. (2009)
(using their cosmological parameters). Again, the crosses are the large-scale
limit of Sheth, Mo \& Tormen (2001) and we only plot our prediction 
(solid lines) down to scale $r=2q$. For $M=10^{11}h^{-1}M_{\odot}$ we also
plot the linearized bias (\ref{bLin}) (lower dashed line), and the nonlinear bias
(\ref{b2M}) where we set $s=r$ (upper dot-dashed line) or we use 
Eq.(\ref{srLin}) (dotted line).
As noticed in Reed et al. (2009), the scale-dependence is much steeper than the
one found at small redshifts and it is not consistent with the fits obtained at
low $z$ in Hamana et al. (2001) or Diaferio et al. (2003).
This was interpreted as a breakdown of universality for massive halos at high
redshift by Reed et al. (2009).
However, we can see that our prediction (\ref{b2M})-(\ref{sr})
agrees reasonably well with their numerical results. 
Therefore, the change of behavior of the bias $b(r)$ between the two regimes
studied in Figs.~\ref{figbiasr_z0} and \ref{figbiasr_z10} can be understood
from the standard picture of massive halos arising from rare overdensities
in the initial (linear) Gaussian density field, by using the same
theoretical prediction (\ref{b2M})-(\ref{sr}) that applies to any $z$.
We can note that the linearized bias (\ref{bLin}),
or the approximation $s=r$, show strong deviations in this regime and
disagree with the simulations. Therefore, one should use the nonlinear bias
(\ref{b2M})-(\ref{sr}) (but using Eq.(\ref{srLin}) gives similar results)
and one cannot neglect the correction due to the Lagrangian to Eulerian
mapping that is associated with $s\mapsto r$.

\begin{figure}[htb]
\begin{center}
\epsfxsize=9 cm \epsfysize=7 cm {\epsfbox{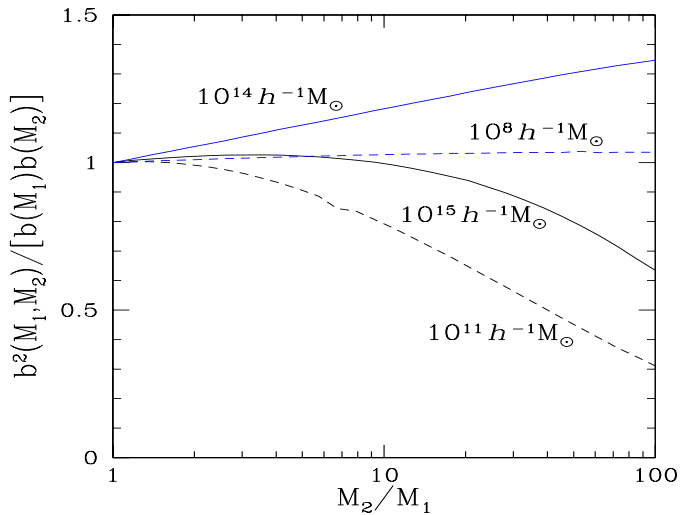}}
\end{center}
\caption{The ratio $b^2(M_1,M_2)/[b(M_1)b(M_2)]$, from Eqs.(\ref{b2M1M2})
and (\ref{b2M}), at fixed geometrical mean $M=\sqrt{M_1M_2}$.
Solid lines are for distance and redshift $(r,z)=(50h^{-1}{\rm Mpc},0)$,
whereas dashed lines are for $(r,z)=(3h^{-1}{\rm Mpc},10)$.
Curves are labelled by their fixed geometrical mean $M$.} 
\label{figbiasom}
\end{figure}

On the other hand, the comparison
with the result from (\ref{bLin}) shows that the steep dependence
on scale is due to the nonlinear term in (\ref{b2M}), i.e. keeping the
exponential factor. Indeed, as noticed below Eq.(\ref{b2M}), and in
Politzer \& Wise (1984), the derivation of the bias presented above only
assumes a large separation between very massive (rare) halos, that is,
$\sigma^2_{q,q}(s)\ll \sigma^2_q$ and $\delta_L/\sigma_q \gg 1$.
Therefore, for sufficiently massive objects (that correspond to a large bias),
the variance $\sigma^2_q$ can be small enough to make the exponent in
Eq.(\ref{b2M}) of order unity or larger. Then, one needs to keep the nonlinear
expression (\ref{b2M}) rather than expanding the exponential as in
Eq.(\ref{bLin}). As noticed below Eq.(\ref{b2M}), this implies a nonlinear
biasing scheme as the halo correlation is not proportional to the matter
correlation but shows a steeper scale dependence. Within the local bias
framework (Fry \& Gaztanaga 1993; Mo et al. 1997), this could be interpreted
as non-zero higher order bias parameters $b_i$ in the expansion
$\delta_{\rm halo}(M)= \sum_i b_i(M) \delta^i_R$. However, the bias
(\ref{b2M}) cannot be exactly reduced to such a model (but one could certainly
derive within such a framework a good approximation to Eq.(\ref{b2M}),
restricted to some larger scale $R$, by using Eq.(\ref{srLin}), writing the
correlations $\sigma^2_{q,0}(r)$ and $\sigma^2_{q,q}(r)$ in terms of
$\sigma^2_{0,0}(r)$, expanding over $\sigma^2_{0,0}(r)$ and finally smoothing
over the external scale $R$ of interest).

We should stress here that our prediction
(\ref{b2M})-(\ref{sr}) has no free parameter. This can lead to a slightly larger
inaccuracy as compared with fits to simulations in the regime where the
latter have been tested, as in Fig.~\ref{figbiasr_z0}, but this improves the
robustness of the predictions as one consider other regimes (e.g. other
cosmological parameters or other redshifts as in Fig.~\ref{figbiasr_z10}).
Therefore, we think that Eqs.(\ref{b2M})-(\ref{sr}) could provide
a useful alternative to current fitting formulae, as they can be readily 
applied to any set of cosmological parameters or redshifts.

Finally, we show in Fig.~\ref{figbiasom} the bias ratio 
$b^2(M_1,M_2)/[b(M_1)b(M_2)]$, as a function of the mass ratio $M_2/M_1$,
at fixed geometrical mean $M=\sqrt{M_1M_2}$. We consider several mass scales
$M$, at distance and redshift $(r,z)=(50h^{-1}{\rm Mpc},0)$ (solid lines)
and $(r,z)=(3h^{-1}{\rm Mpc},10)$ (dashed lines). This shows that making
the factorized approximation, $b^2(M_1,M_2) \simeq b(M_1)b(M_2)$,
can lead to an error of up to $50\%$ for a mass ratio $M_2/M_1 \sim 100$.
Therefore, it is best to use the full Eq.(\ref{b2M1M2}).

\section{Conclusion}
\label{Conclusion}

We have pointed out in this article that the large-mass exponential tail
of the mass function of collapsed halos is exactly known, provided halos
are defined as spherical overdensities above a nonlinear density contrast
threshold $\delta$ (i.e. using a spherical overdensity algorithm in terms of
numerical simulations). This arises from the fact that massive rare
events are governed by (almost) spherical fluctuations in the initial (linear)
Gaussian density field (if one does not explicitly breaks statistical isotropy
by looking for non-spherical quantities).
This is most easily seen from a steepest-descent
approach, which becomes asymptotically exact in the large-scale limit,
applied to the action $\cS$ associated with the probability distribution
of the nonlinear density contrast within spherical cells.
This result holds for any nonlinear threshold $\delta$ used to define halos,
provided it is below an upper bound $\delta_+$ that marks the point where
shell-crossing comes into play. For a standard $\Lambda$CDM cosmology,
$\delta_+$ typically grows from $200$ to $600$ as one goes from
$10^{15}h^{-1} M_{\odot}$ to $10^{11}h^{-1} M_{\odot}$ (which also corresponds
to increasing redshift). This dependence on mass is due to the change of
slope of the linear power spectrum with scale.

We have also noted that in two similar systems, the one-dimensional adhesion
models with Brownian or white-noise initial (linear) velocity,
the same method can be used for both the density distribution $\cP(\delta_r)$
and the mass function $n(M)$, and one can check that this yields rare-event
tails that agree with the exact distributions, which can be derived
by other techniques (Valageas 2009a,b). 

Therefore, defining collapsed halos by a threshold $\delta=200$ to follow
the common practice, the large-mass tail of the halo mass function 
is of the form $e^{-\delta_L^2/(2\sigma^2(M))}$, up to subleading prefactors
such as power laws, where $\delta_L=\cF^{-1}(\delta)$ is the linear density
contrast associated to $\delta$ through the spherical collapse dynamics.
In particular, we obtain $\delta_L\simeq 1.59$ for $\delta=200$.
We checked that this value, which is slightly lower than the
commonly used value of $\delta_c=1.686$ associated with complete collapse,
gives a good match with numerical simulations (at large masses) when we simply
use the Press-Schechter functional form. We also give a fitting formula,
which obeys this exact exponential cutoff, that agrees with simulations over
all mass scales. 

We suggest that halos should be defined by such a nonlinear density threshold
(i.e. friends-of-friends algorithms are not so clearly related to theoretical
computations) and fits to numerical simulations should use this exact
exponential tail, rather than treating $\delta_L$ as a free parameter.
This would avoid introducing unnecessary degeneracies between fitting parameters
and it would make the fits more robust.

Next, we have briefly recalled that in the large-mass limit the outer density
profile of collapsed halos is given by the radial profile of the relevant
spherical saddle-point. This applies to radii beyond the density threshold
$\delta_+$, where shell-crossing comes into play. In agreement with numerical
simulations, for rare massive halos this separates an outer region dominated
by a radial flow from an inner region where virialization takes place
and a strong transverse velocity dispersion quickly builds up. We have recalled
that this can be explained from a strong radial-orbit instability, which implies
that infinitesimal deviations from spherical symmetry are sufficient to
govern the dynamics.

Finally, following the approach of Kaiser (1984), we have obtained an analytical
formula for the bias of massive halos that improves the match with numerical
simulations. In particular, it captures the steepening of the scale dependence
that is observed for large-mass halos at higher redshifts. This requires
keeping the bias in its nonlinear form and taking care of the
Lagrangian-Eulerian mapping. We also note that using a factorization
approximation, $b^2(M_1,M_2) \simeq b(M_1)b(M_2)$, may lead to non-negligible
inaccuracies. We stress that this analytical estimate of the bias contains
no free parameter. Although this can yield a match to numerical simulations
that is not as good as fitting formulae derived from the same set of simulations,
it provides a more robust prediction for general cases, as shown by the
good agreement obtained at both low and high redshifts ($z=0$ and $z=10$),
whereas published fitting formulae cannot reproduce both cases.
We think this makes such a model useful for cosmological purposes, where it is
desirable to have versatile analytical estimates that follow the correct trends
as one varies cosmological parameters or redshifts.
In particular, the scale-dependence of the bias of massive halos has recently
been proposed as a test of primordial non-Gaussianity (e.g., Dalal et al. 2008),
which requires robust theoretical models.

Our results are exact (for the mass function) or are expected to provide
a good approximation (for the bias) in the limit of rare massive halos.
However, this remains of interest as large-mass tails are also the most
sensitive to cosmological parameters (e.g., through the linear growth factor
and the primordial power spectrum), thanks to their steep dependence on mass
or scale. Moreover, we think that more general fitting formulae (such as the
one we provide for the mass function) should follow such theoretical predictions
in their relevant limits, so as to reduce the number of free parameters and
improve their robustness.

\end{document}